\documentclass[]{uiophd}
\usepackage[utf8]{inputenc}         
\usepackage{commath}
\usepackage{breqn}
\usepackage{amsmath,bm}
\usepackage{natbib}
\usepackage{subfigure}

\title{An investigation into the interaction between waves and ice}
\author{Jean Rabault}

\begin{document}
\frontmatter
\maketitle
\tableofcontents

\mainmatter

\chapter*{Abstract}
\addcontentsline{toc}{chapter}{Preface}

This thesis is submitted in the partial fulfillment of the requirements for the degree of Doctor of Philosophy at the University of Oslo. It represents work that has been carried out between 2015 and 2018, under the supervision of Pr. Atle Jensen, Dr. Graig Sutherland, and Dr. Kai H. Christensen, in collaboration with Pr. Aleksey Marchenko and Pr. Brian Ward. The work presented was carried at the University of Oslo and the University Center in Svalbard. Financial support for the work was provided by the Norwegian Research Council under the Petromaks 2 scheme, through the project WOICE (Experiments on Waves in Oil and Ice), NFR Grant number 233901.

The thesis consists of an introduction, and a selection of 7 publications. The introduction presents the scientific context in which the work was undertaken, the methodology used, the results obtained, as well as some personal thoughts about unsuccessful directions encountered during the project and possible future work. I certify that this dissertation is mine and that the results presented are the result of the work of our research group, to which I brought significant contribution.

~ \\
~ \\

\noindent Oslo, March 2018 \\
\noindent Jean Rabault \\

\chapter*{Acknowledgements}
\addcontentsline{toc}{chapter}{Acknowledgments}

I want to express my deepest gratitude towards my PhD advisors, Pr. Atle Jensen, Dr. Graig Sutherland, and Dr. Kai H. Christensen. Their help and guidance have been critical to the work accomplished. In addition, I have received precious help from a number of people in the course of the project. Pr. Aleksey Marchenko has been a key player for this project both during fieldwork and laboratory experiments on Svalbard. The participation of Pr. Brian Ward was critical for collection of the data obtained during the first field measurements. Finally, MSc. Trygve Halsne has paved the way for several of the laboratory experiments performed at the University of Oslo through his Master Thesis. Moreover, none of this work would have been possible without the help of our Laboratory Engineer, MSc. Olav Gundersen, whose help, technical competence, and good spirit is key to all the successes achieved throughout this project. In the same spirit, I want to express special thanks to Dr. Jostein Kolaas, who is at the origin of much of the software and scripts used when performing PIV. I also want to thank Terje Kvernes and Lucy Karpen, for their kind help, and interesting discussions covering a broad range of computer network, hardware, and other computer science related topics.

Finally, I want to thank all the people I have met either during or alongside work in the last $4$ years, who have made my life in Norway a nice moment, as well as my family for their continuous support.

\chapter{Introduction}

\section{Polar regions: key areas under rapid evolution}

\begin{figure}
  \begin{center}
    \includegraphics[width=.65\textwidth]{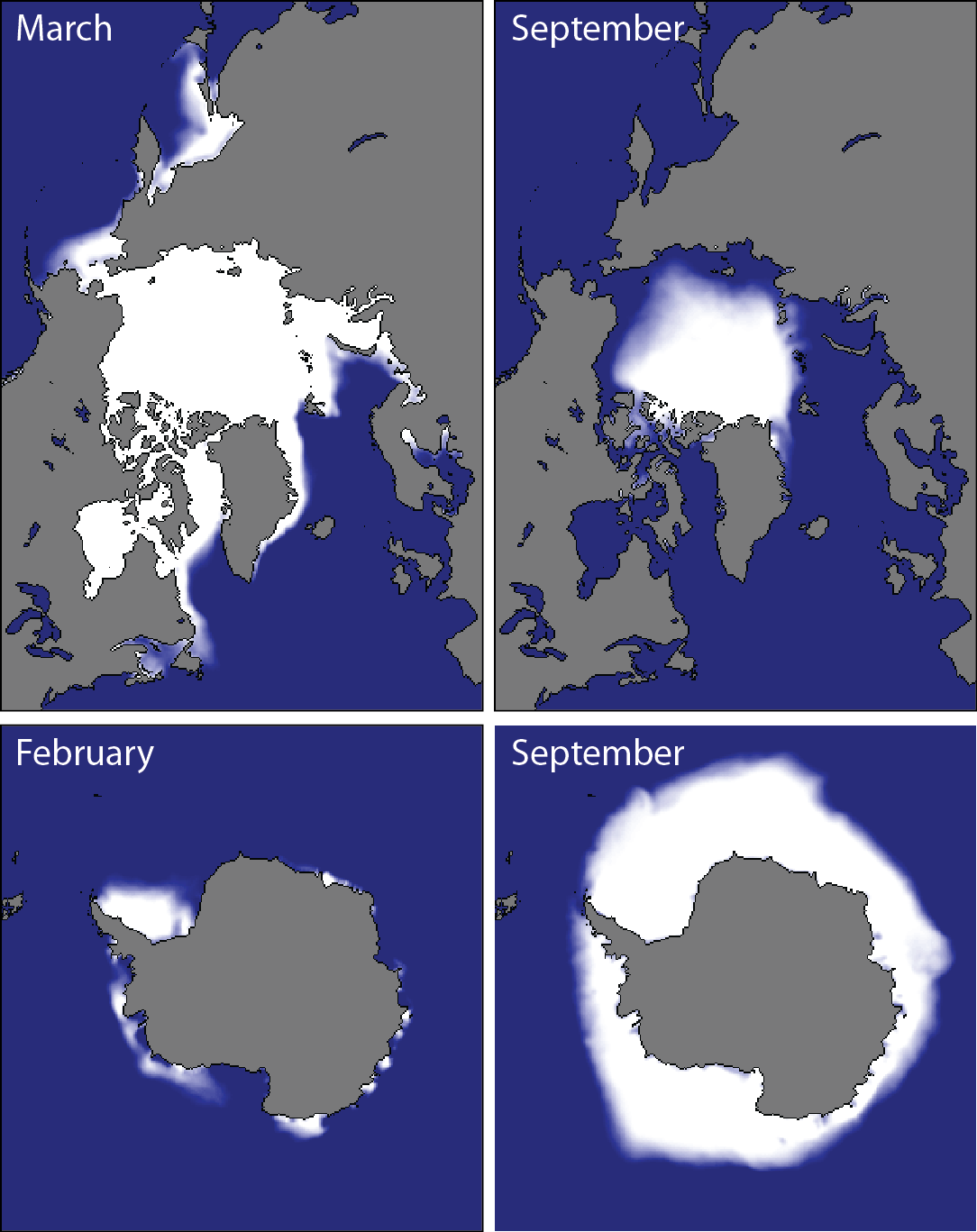}
    \caption{\label{polar_regions}Average Arctic and Antarctic sea ice extent from the period 1981-2010 based on microwave satellite data, at the approximate seasonal maximum and minimum levels for each of the polar regions. Reproduced from information released by the National Snow and Ice Data Center (NSIDC), University of Colorado, Boulder.}
  \end{center}
\end{figure}

Sea ice is a major feature of the polar environments, both in the Arctic and Antarctic. Sea ice extent reaches an annual maximum of about 14 million square kilometers in the Arctic, and 18 million square kilometers in the Antarctic (see Fig. \ref{polar_regions}). In average throughout the year, sea ice covers a total of 25 millions square kilometers, which is about 7 \% of the world's oceans surface \citep{parkinson1997earth, council2007polar}. Sea ice is both a central element of the polar ecosystems and a key to the global climate and ocean circulation.

In the last couple of decades, significant decline in the Arctic sea ice extent has been observed \citep{Feltham20140171}. While accurately predicting the pace at which sea ice decline will occur remains challenging \citep{GRL:GRL50316}, the reduction of sea ice extent in the Arctic cannot be denied. Predicting the evolution of sea ice extent is made challenging by the number of contributing feedback mechanisms, and the complex physics involved. Some feedback mechanisms involved in sea ice decline have been well known for a long time, such as the Albedo difference between open sea and sea ice: while ocean surface without sea ice reflects about 6 to 7 \% of the incoming solar energy flux, sea ice covered by new snow can reflect as much as 85 \% of the incoming energy \citep{1748-9326-10-6-064001}. Other feedback mechanisms have been discovered more recently, and may also play an important role in particular in the Arctic. One such mechanism is the consequence of ice breaking by surface gravity waves. It has been shown from satellite data that reduced sea-ice extent, by increasing the fetch available for wave development, has lead to higher wave activity and wave amplitude in the Arctic basin, which in turn increases ice breakup and melting \citep{GRL:GRL51656, KohoutScienceBreakupWaves}.

While sea ice is important for the global climate, it is also critical to human activities in the polar regions. Changes in the climate alongside technological developments are creating new opportunities for human activities in the Arctic, including sustainable development of resource-based industries, fishing, tourism, and faster shipping routes between Europe and Asia. Improved scientific understanding of the Arctic environment, especially sea ice dynamics that are largely influenced by the incoming wave field, are necessary to help produce increased value in polar regions, while doing so in a safer and more environmental-friendly way \citep{KaiReport, Pfirman1995129, Rigor199789}.

Therefore, there is motivation to develop a better understanding of sea ice and its interaction with waves, both for improving our understanding of the global climate system, protecting the local ecosystems, and enabling larger scale human activities in the polar regions.

\section{Sea ice and waves: a brief overview of the literature}

\begin{figure}
  \begin{center}
    \subfigure[]{\includegraphics[width=.45\textwidth, height=.2\textheight]{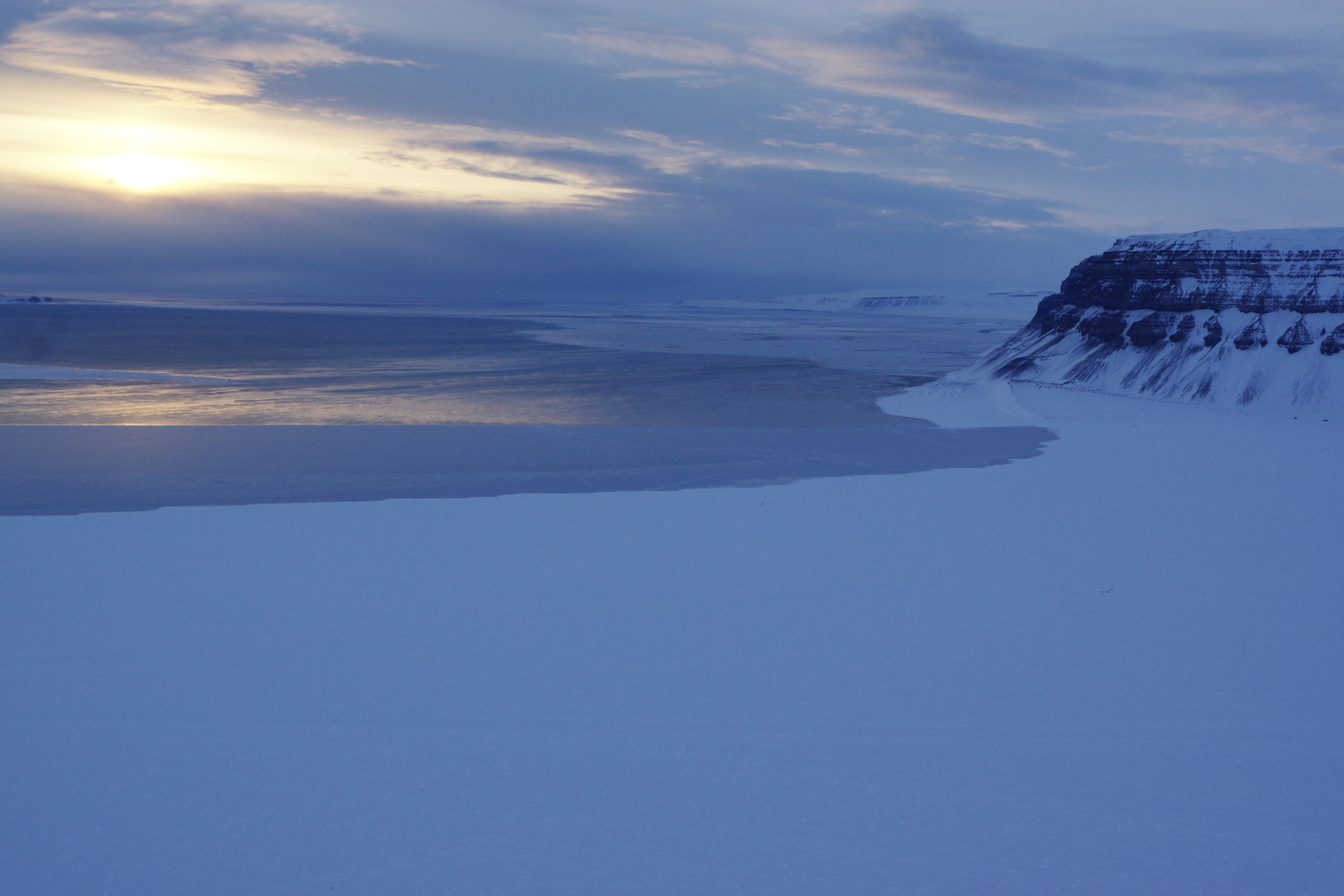}}
    \subfigure[]{\includegraphics[width=.45\textwidth, height=.2\textheight]{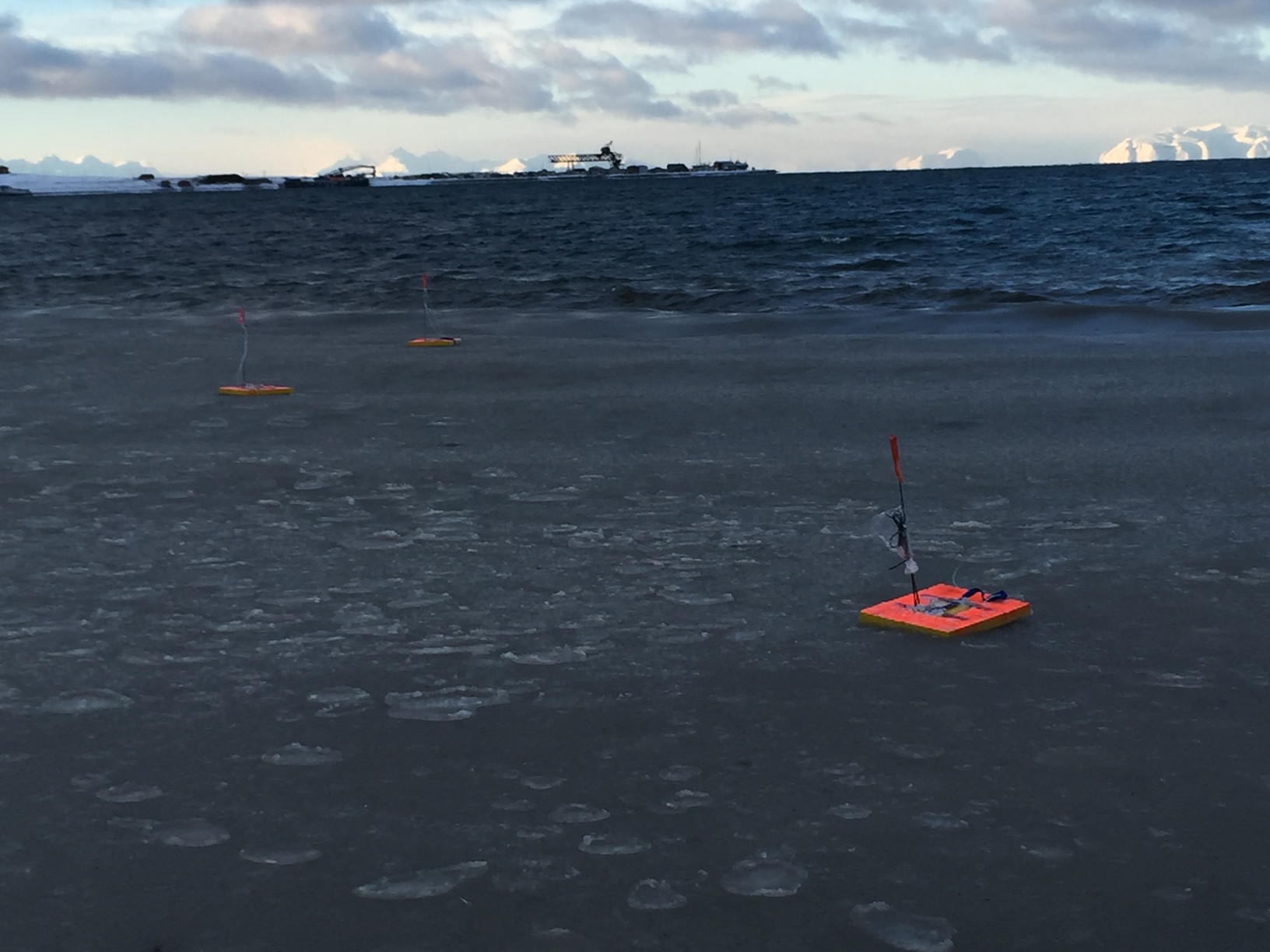}} \\
    \vspace{0.1cm}
    \subfigure[]{\includegraphics[width=.45\textwidth, height=.2\textheight]{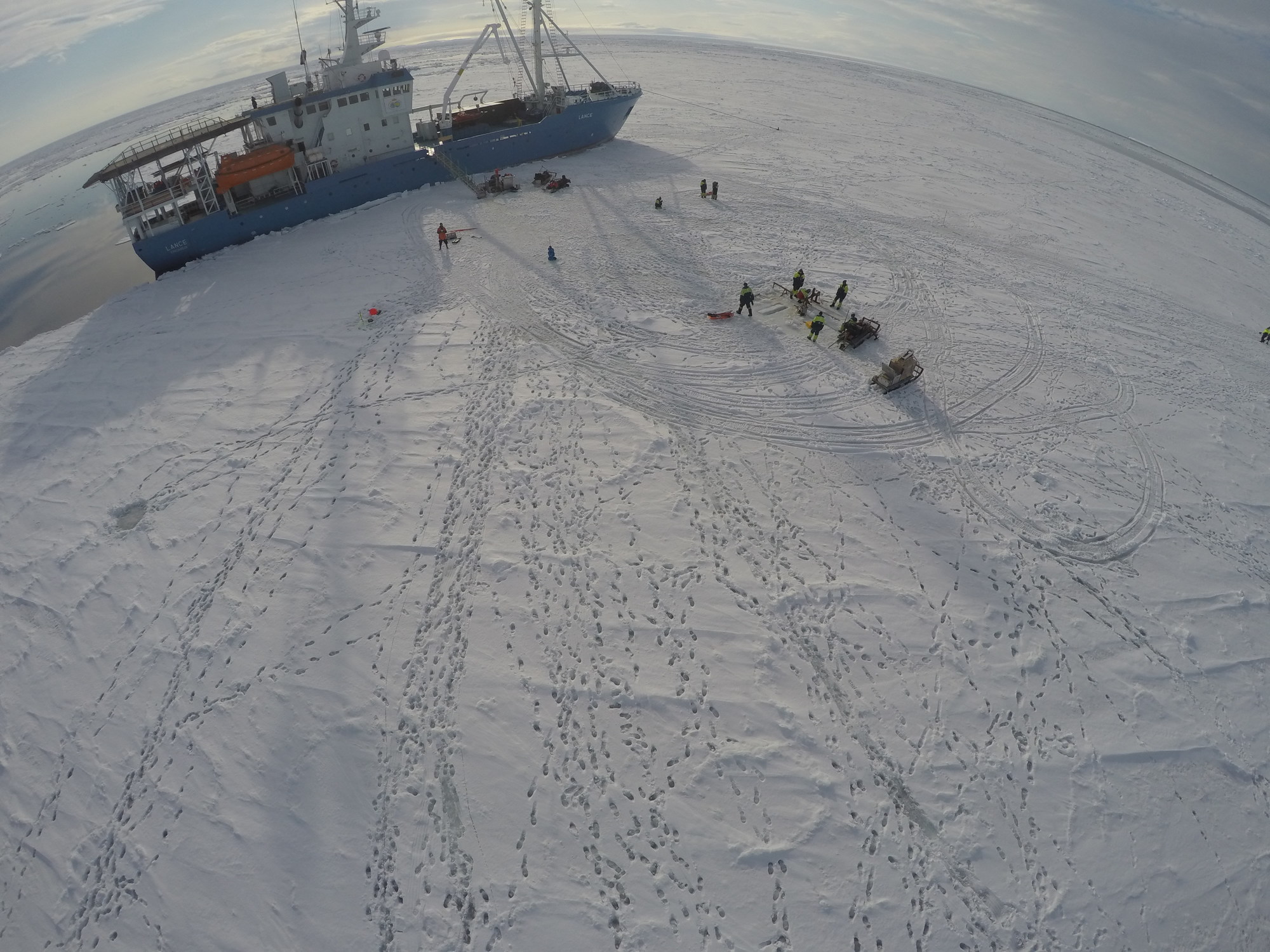}}
    \subfigure[]{\includegraphics[width=.45\textwidth, height=.2\textheight]{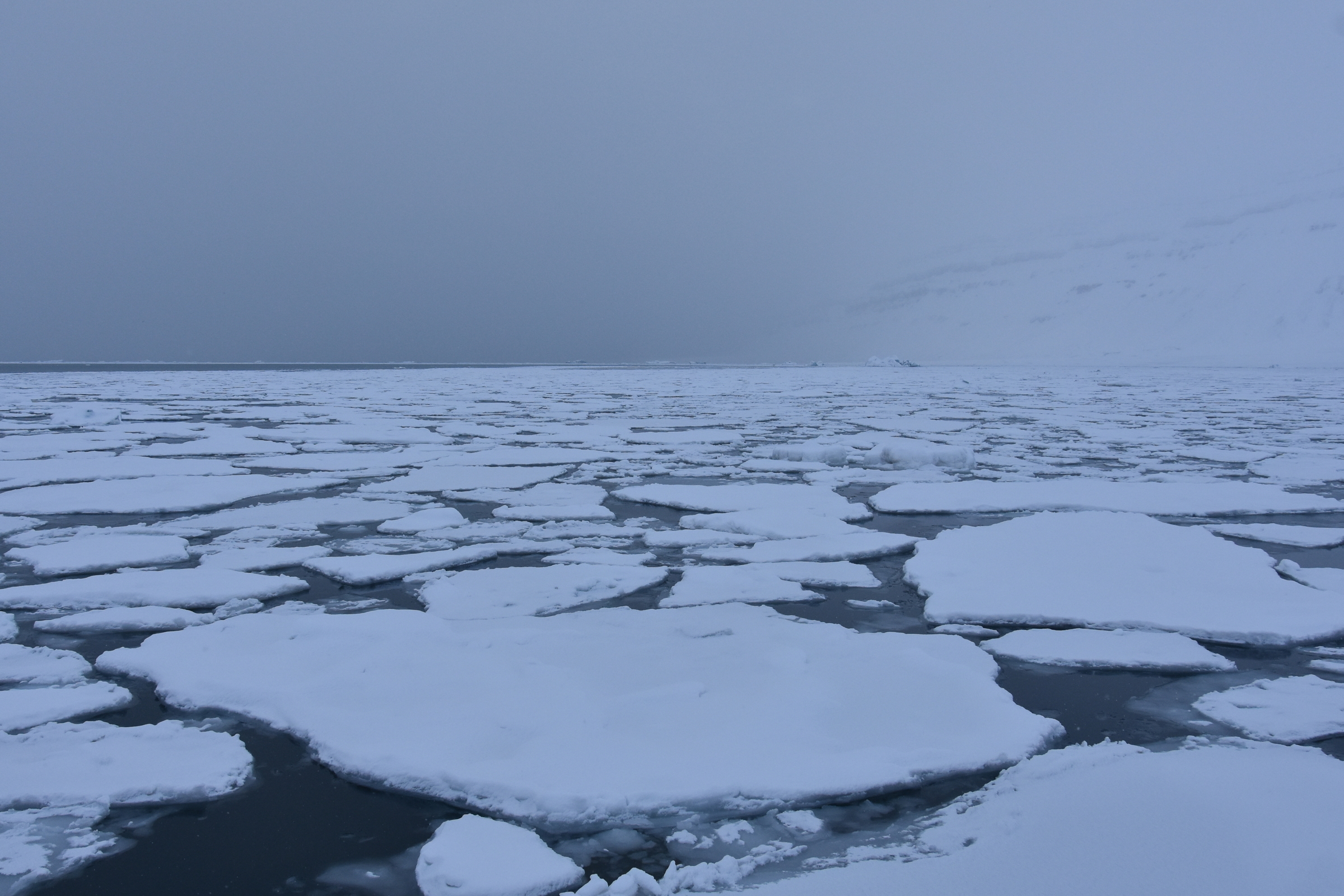}} \\
    \vspace{0.1cm}
    \subfigure[]{\includegraphics[width=.45\textwidth, height=.2\textheight]{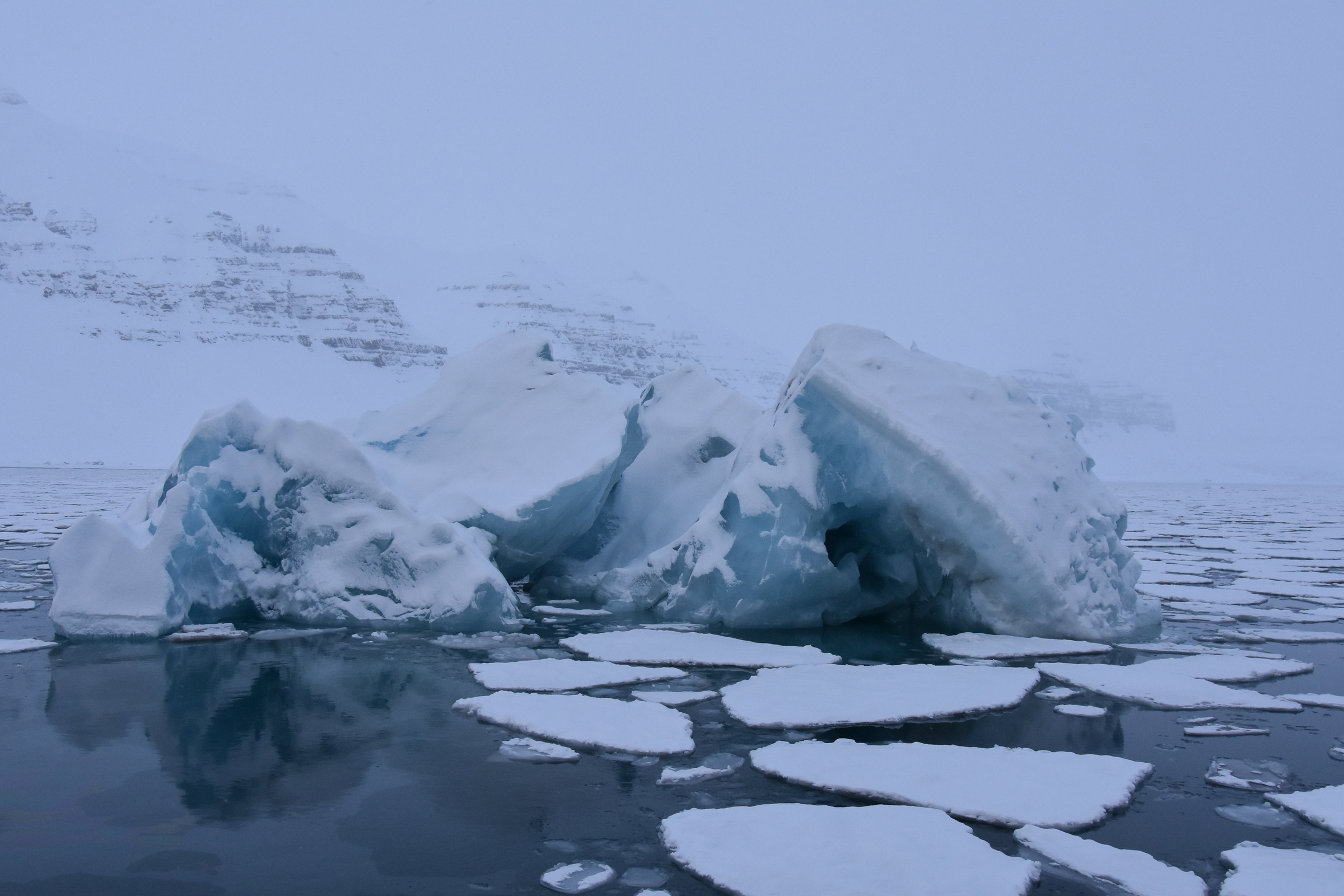}}
    \caption{\label{ice_diversity} Illustration of the diversity of ice conditions encountered during the course of the WOICE the project. Credits: WOICE members (Atle Jensen and Jean Rabault, UiO; Aleksey Marchenko, UNIS; Graig Sutherland, MET), and Sebastian Sikora (UNIS logistics). (a): landfast ice and newly formed ice fringes (Tempelfjorden, 2015), (b): grease ice slick close to the shore (Adventfjorden, 2016), (c): large ice floe in the Barents sea (East of Svalbard, 2016), (d): broken floes (Tempelfjorden, 2017), (e): iceberg released from glacier calving (Tempelfjorden, 2017). Each of those forms of ice have different physical properties that influence wave propagation and damping, in addition to providing a variety of opportunities for the local species, as well as challenges for human activities.}
  \end{center}
\end{figure}

The study of sea ice is made complex by the great variety of ice conditions that are observed in nature. Ice comes in a great variety of shapes, thicknesses and mechanical properties, from grease ice (behaving similarly to a floating layer of highly viscous fluid) to land-fast ice (behaving as a thin elastic plate), through pack ice, large floes, and icebergs \citep{SquireOOWASI, NewyearLab1, Squire2007110, KaiReport, JGRC:JGRC21649}. This is illustrated by images taken throughout the course of the project (Fig. \ref{ice_diversity}). Each kind of ice exhibits different physical properties, including density, mechanical resistance, brittleness, Young modulus, and the amount of trapped air, salt and brine. The combination of these different types of ice is important for the existence of a large Arctic ice pack. For example, grease ice and broken ice floes present in the Marginal Ice Zone (MIZ, the first ice infested area encountered from the open ocean) play an important role in damping high frequency waves, that would otherwise lead to fast breaking of the inner, continuous sea ice.

The following sections present a brief overview of the current understanding of wave-ice interaction that is available through both experimental data and theoretical analysis published in the literature.

\subsection{Fieldwork and experiments of wave-ice interaction}

Despite a long tradition of studying waves in ice \citep{1023072369499, ewing1934propagation}, field measurements are still relatively sparse when considering the large area and diversity of the polar regions. This can in part be explained by the dangerous conditions in which waves in ice should be measured, challenges in the deployment of sensors, and the costs associated with the development of instruments that can work in the polar conditions \citep{Squire2007110}.

When considering wave-ice interaction, a number of quantities should ideally be measured. The most important of these include the wave dispersion relation, wave spectrum, wave directional spectrum (which, by contrast to wave spectrum, contains not only information about the energy content at each frequency, but also information about the direction of propagation and angular spread), ice quality and thickness, wave damping, and water eddy viscosity. Measuring the wave dispersion relation is important from a theoretical perspective, and has also been suggested as a way to indirectly estimate the ice quality and thickness \citep{marsan2012sea}. Indeed, ice thickness influences the dispersion relation \citep{1023072369499} and one can, therefore, obtain an estimation of the ice thickness from measurements of wave propagation. Similarly, the wave spectrum contains important information about the incoming wave energy, and measurements at different locations can be used to compute the wave damping as a function of frequency. This, in turn, provides important information for the calibration of wave in ice models. Finally, measuring the wave directional spectrum allows for the gathering of additional information about wave propagation. In particular, it has been suggested that the wave directional spectrum can be a way of discriminating which mechanism is dominating the wave attenuation, either non-conservative dissipation or conservative directional redistribution \citep{ardhuin2016ocean}.

In the literature, two main categories of methods are presented for performing measurements of the propagation of waves in ice. The first category relies on instruments that need to be physically present at the location of measurement, either on or under the ice. These instruments can rely on a variety of sensors such as seismometers \citep{marsan2012sea}, pressure sensors \citep{marchenko2013measurements}, shear probes \citep{mcphee1994turbulent}, accelerometers, Inertial Motion Units (IMUs) \citep{kohout2015device, GRL:GRL51947}, Acoustic Doppler velocimeters (ADVs), and Acoustic Doppler current profiler (ADCPs) \citep{1506330}. Such sensors allow a direct measurement of the ice or water motion, and the data must be stored locally and retrieved physically at the location of the instrument. Alternatively, it may be sent through Iridium or another satellite communication means owing to the remoteness of the polar regions. Using such instruments makes it relatively easy to compute simple quantities such as wave spectrum, wave damping and the water eddy viscosity \citep{mcphee1994turbulent, GRL:GRL51947, GRL:GRL53001}. In addition, one can retrieve the wave number and, therefore, the dispersion relation from a correlation analysis between adjacent instruments \citep{fox_haskell_2001}, even though such measurements remain sparse \citep{Squire2007110}.

By contrast, another category of measurements relies on instruments that do not need to be physically present on the ice, in particular Synthetic Aperture Radars (SARs, \citet{liu1991wave, JGRC:JGRC22637}) and lasers (LIDAR, \citet{wadhams1975airborne, GRL:GRL54296}). These methods have been deployed from both planes and satellites, and offer several advantages, the most obvious one being that they do not require physical access to the ice, which can be both difficult and dangerous. SAR images obtained from satellites such as the two Sentinel 1 are especially convenient, as these follow a polar orbit. As a consequence, the polar regions are very well covered and large datasets are freely available. After processing, SAR images allow the computation of important quantities such as the quality of the ice cover \citep{articlesarseaice}, or the wave 2D spatial spectrum \citep{JGRC:JGRC22637}. However, some limitations are also present when using SAR satellite images. In particular, usually SAR images do not include time resolved information, making it challenging to retrieve the frequency data necessary for the computation of the dispersion relation \citep{JGRC:JGRC22637}.

In parallel to field measurements, laboratory experiments have been performed at least since the early 1980s. These include both small scale measurements in wave tanks typically a few meters long and half a meter deep kept in a cold room \citep{Martin1981, NewyearLab1}, and measurements in large scale environmental wave flumes \citep{Wang201090, Zhao201571}. Most commonly, the waves are measured using pressure sensors located on the wall of the tank \citep{NewyearLab1, Wang201090}, a few tens of centimeters under the surface of the water. This presents some challenges, as pressure is a quantity that can be difficult to measure accurately (since the pressure variations due to waves can often be small compared to the mean pressure value at a given measurement depth), and the pressure disturbance created by the waves is exponentially attenuated with depth. In addition, performing measurements on the side wall of the wavetank may induce some systematic bias if the wave field is 3D. All those inconveniences can be avoided using ultrasonic gauges, but these have not be commonly used in the past. Measurements of wave elevation can be used to obtain the wave damping, as well as the wavelength (and, therefore, dispersion relation). Wavelength is obtained through a correlation analysis granted that several measurements are performed within a distance of typically one half wave length. Therefore, by performing measurements for a series of monochromatic waves of varying frequency, one can explore all aspects of wave propagation in the laboratory.

Unfortunately, while wave propagation has been investigated in several wave-tank experiments, very little data have been collected regarding the motion of the water, such as mean currents and eddy viscosity. This is the consequence of the higher cost, complexity, and operational challenges implied by instruments that can perform such measurements. To the knowledge of the author, only the study of \citet{Martin1981} has discussed such aspects, qualitatively and based on visual observations.

\subsection{Theoretical understanding}

One of the aims of experimental data is to provide a basis for the development and calibration of models describing the propagation of waves in ice. The ability to understand and model both wave damping and wavelength is important for a number of practical applications. For example, predictions of the wave-field in the polar regions, which is critical for safety of operations in the arctic, weather forecasts, and sea ice forecasts, requires an estimate of the spatial damping of waves \citep{KaiReport}.

When studying the propagation of waves in ice infested regions, one usually considers a single 2D harmonic wave mode having a frequency $f = \omega / (2 \pi)$, wave number $k = 2 \pi / \lambda$, and surface elevation $\eta(x, t) = a \cos(k x - \omega t)$, where $a$ is the wave amplitude, $t$ the time and $x$ the direction of wave propagation. Owing to the superposition principle, any wave state can be reconstructed from a combination of such harmonic modes, at least as far as linear effects are the dominant mechanism so that different wave modes do not interfere with each other and can be considered individually.

The first theoretical description of wave propagation in ice-covered sea was presented by \citet{1023072369499}, and describes the propagation of waves under a continuous, solid ice cover behaving as a thin elastic plate. Such description of wave propagation under a continuous solid ice cover has proven successful and was further refined by a number of authors \citep{SQUIRE1993219, JGRJGR14705, AnalysisPolarsten, SquireOOWASI}, to include for example the effect of compression stress of the ice cover. The main idea of the model consists in conducting the classical derivation for water waves relying on the potential flow theory, while changing the boundary condition at the water surface to take into account the elastic effect of the ice. This model describes the effect of the ice on the dispersion relation, but does not consider energy dissipation and damping. Stream function models can also be used to describe the propagation of waves in ice, but are generally encountered for models that aim at describing wave damping. Considering a 2D monochromatic wave traveling in the positive $x$ direction, with the vertical direction $z$ pointing upwards, $z=0$ being the mean water surface (covered by ice), $z = -H$ the water bottom, one gets the following system of equations in the frame of potential flow theory:

\begin{itemize}
  \item Equations for velocity components:
  \begin{equation}
    u_x = \frac{\partial \phi}{\partial x}, ~ u_z = \frac{\partial \phi}{\partial z}
    \label{potential_flow}
  \end{equation}

  \item 2D continuity equation in the water occupied volume:
  \begin{equation}
    \frac{\partial^2 \phi}{\partial x^2} + \frac{\partial^2 \phi}{\partial z^2} = 0
    \label{continuity}
  \end{equation}

  \item Free slip boundary condition at the water bottom, $z = -H$, in the case of potential flow:
  \begin{equation}
    \frac{\partial \phi}{\partial z} = 0
    \label{BC_bottom}
  \end{equation}

  \item Kinematic equation for the water / ice interface, $z = \eta(x, t)$:
  \begin{equation}
    \frac{\partial \eta}{\partial t} + \boldsymbol{u} \cdot \nabla \eta = \frac{\partial \phi}{\partial z}
    \label{BC_kinematic}
  \end{equation}

  \item Pressure balance at the water / ice interface, $z = \eta(x, t)$:
  \begin{equation}
    - \rho_w \left( \frac{\partial \phi}{\partial t} + g \eta \right) = \left( \frac{E h^3}{12 (1 - \nu^2)} \frac{\partial^4}{\partial x^4} + h P \frac{\partial^2}{\partial x^2} + h \rho_i \frac{\partial^2}{\partial t^2} \right) \eta(x, t)
    \label{BC_thin_plate}
  \end{equation}
\end{itemize}

\noindent where $\boldsymbol{u}(x, z) = (u_x (x, z), u_z (x, z))$ is the 2D wave velocity field, $\phi (x, z)$ the associated velocity potential, $g$ the acceleration of gravity, $\rho_i$ the volumetric mass density of the ice, $\rho_w$ the volumetric mass density of the water, $h$ the thickness of the elastic ice cover, $E$ its elastic modulus, $\nu$ its Poisson coefficient, and $P$ the compression stress. The left hand side of Eqn. (\ref{BC_thin_plate}) is the pressure in the water as described from the Bernoulli equation, while its right hand side comes from the application of thin plate theory to the elastic ice cover.

Linearizing Eqns. (\ref{BC_kinematic}) and (\ref{BC_thin_plate}) around the state of rest, and keeping only first order terms, results in a system of equations that can be solved in a way similar to the usual linear water waves. The dispersion relation obtained is then of the form:

\begin{equation}
  \omega^2 = \frac{gk+Bk^5-Qk^3}{\coth(kH)+kM},
  \label{DispRelIW}
\end{equation}

\noindent where the following reduced parameters were used:

\begin{equation}
  B = \frac{E h^3}{12 \rho_w (1-\nu^2)}, \quad Q = \frac{P h}{\rho_w} , \quad M = \frac{\rho_i h}{\rho_w}.
  \label{ValueCoeffs}
\end{equation}

The dispersion relation Eqn. (\ref{DispRelIW}) and the phase and group velocities, $c_p = \omega / k$ and $c_g = \partial \omega / \partial k$, can then be obtained numerically. The predictions for wavelength compares well with data from solid, continuous ice cover, even if relatively few such studies are available in the literature \citep{JGRC:JGRC3357}.

A few observations can be made using Eqns. (\ref{DispRelIW}, \ref{ValueCoeffs}). First, in the case without ice cover ($B = Q = M = 0$), the usual water waves dispersion relation is obtained: $\omega^2 = g k \tanh(kh)$. In the case that there are no compression and flexural effects (for example, because many cracks are present in the ice cover), the simpler mass loading model is recovered. One should note that the ice thickness has a very drastic effect on the dispersion relation, as $B \propto h^3$, and flexural rigidity appears in the dispersion relation through a term in the form $B k^5$.

While theory is relatively successful at describing wave propagation under a continuous ice cover, the problems of describing wave attenuation in all types of ice and wave propagation in complex ice conditions are more arduous. In the case of a continuous solid ice cover, attenuation can take place either in the ice itself, due to inelastic deformation of the ice cover, brine migration, creep, plasticity, and other complex mechanisms \citep{collins2016wave, marchenko2017influence, JGRC:JGRC11467}, or in the water under the ice due to the effect of viscosity on the shear generated by the no-slip boundary condition at the ice interface \citep{lamb1932hydrodynamics, WeberArticle}. The first category of phenomena, i.e. dissipation in the continuous ice itself, is difficult to describe and predict in details as the properties of the ice are very dependent on a number of parameters such as ice age, brine content, and temperature. The second category of phenomena is also challenging to predict due to the influence of parameters such as the roughness of the interface between the water and the ice sheet, and the influence of the eddy viscosity level in the water on dissipation \citep{DeCarolis2002399}. The situation is even more complicated when complex ice such as ridged inhomogeneous pack ice, pancake ice or broken ice floes are present: in this case, the geometry and spatial variability of the ice add complexity to the formulation of the problem and new phenomena, such as wave diffraction and reflection, become important.

When studying wave damping by sea ice due to viscous effects, one generally considers that the wave amplitude is exponentially damped across the direction of propagation:

\begin{equation}
\frac{\partial a}{\partial x} = - \alpha a,
\end{equation}

\noindent where $x$ the distance into the ice-covered region, $a$ the wave amplitude, and $\alpha$ the spatial decay coefficient that describes wave attenuation. It should be noted that one can easily convert between spatial and temporal damping using the relation from \citet{gaster1962}:

\begin{equation}
  \frac{\beta}{\alpha} = c_g,
\end{equation}

\noindent where $\beta$ is the temporal attenuation rate, and $c_g$ the group velocity. In the following, I will prefer to refer to spatial damping coefficients.

The first theoretical model describing wave attenuation by grease ice is the one by \citet{WeberArticle}. In this model, the ice is considered so viscous that it does not allow appreciable shear, and the balance between friction and pressure results in a ``creeping motion'' of the ice. This imposes a no-slip boundary condition under the ice, and all the dissipation occurs in the underlying water. This is similar to the solution found by \citet{lamb1932hydrodynamics} (Equation 304.xiii), and the resulting damping can be written as:

\begin{equation}
  \alpha = \frac{1}{2} \nu \gamma k / c_g, \label{weber_equation}
\end{equation}

\noindent where $\nu$ is the effective viscosity, and $\gamma =  \sqrt{\omega / 2 \nu}$ is the inverse boundary layer thickness.

An effective eddy viscosity much higher than the molecular viscosity of water is required in the water layer for the model to be consistent with observations and laboratory experiments. While a crude simplification of reality, this simple model provides good agreement with both laboratory and field data when an empirical fit value for the effective viscosity of the water is used \citep{NewyearLab1, RabaultSutherlandGlaciology}. The use of a high effective viscosity is usually justified by the need to describe the existence of a large range of eddies under the ice that enhance mixing \citep{DeCarolis2002399, GRL:GRL53001}, but in addition it could include some of the effects of the highly viscous grease ice layer that may not be described correctly by the model.

Next, \citet{NewyearLab1} introduced another one-layer model to compare with experimental results in grease ice in a small wave tank facility. By contrast with the model of \citet{WeberArticle}, the model of \citet{NewyearLab1} assumes that, since most of the wave motion is usually concentrated near the free surface, most of the wave motion takes place into the grease ice and, therefore, attenuation can be calculated considering an infinitely deep grease ice layer. This solution is similar to the calculation of \citet{lamb1932hydrodynamics} for waves propagating in a viscous fluid (Equation 301.9), which can be expressed as:

\begin{equation}
  \alpha_{ns} = 2 \nu k^2 / c_g,
  \label{damping_NM97}
\end{equation}

\noindent where there also $\nu$ is taken as a fitting parameter, which this time models grease ice viscous properties.

This second model was later adapted into a two layer model by \citet{TwoLayersModel}, who considered the case of finite depth with inviscid water under the ice layer. This opened the way for the development of a variety of such two-layer models. \citet{DeCarolis2002399} formulated a model in which viscosity is also added to the water under the ice, while \citet{JGRC:JGRC11467} considered a model in which the water layer is inviscid, but the ice layer follows a viscoelastic model (Voigt model). Such models are a better description of reality, as a clear separation between grease ice and water is observed in the experiments, and these two phases have vastly different properties. Better agreement is also observed between these models and laboratory data than with the previously mentioned one layer models \citep{NewyearLab2}, even though the estimation of model quality has often been based on curve fitting and visual impression rather than quantitative metrics (see, for a discussion of this issue, \citet{RabaultSutherlandGlaciology}). These models predict attenuation as the solution of complex equations. Therefore, the wave length and attenuation predicted by two layer models cannot be written using simple functions.

Despite their mathematical sophistication, two layer models are affected by at least two issues. First, while being a better description of reality, the parameters they rely on are in practice obtained from fit to experimental data rather than analysis of the underlying properties of the ice, and as a consequence they differ little from the single layer models on this aspect. Therefore, it is difficult to know whether the better agreement observed with experimental data is due to a better description of the physics, or if it is a mathematical artifact due to more fitting parameters being available. In addition, some of the most recent models \citep{TwoLayersModel,Wang201090} have several issues that were raised by \citet{JGRC:JGRC21350}. In particular, the large number of roots in the dispersion relation of the later model makes it challenging to use, as the selection of the physically relevant propagation mode is made much more challenging than in the case of simpler models. While it was recently proven that the dominant mode can easily be identified in most situations \citep{zhao2017nature}, more work is needed to clarify the situation in the mode swap zone, where both flexural-gravity and the pressure wave have similar wave numbers. As a consequence of these difficulties and the inherent complexity of implementation of two layer models, I have focused my efforts towards comparison of experimental data with one layer models, while making the data available to authors who may wish to compare them with the more sophisticated two layer models.

Finally, another effect of wave-ice interaction that can be expected on the field and in experiments is the creation of mean currents under the ice. Such currents can have an influence on important mechanisms such as pollutants and nutrients distribution. A major reason for induced currents under sea ice could arise from diffusion of the vorticity generated in the water-ice boundary layer due to the no-slip boundary condition. This can also be seen as a consequence of wave damping and conservation of momentum \citep{Martin1981}: as waves carry with themselves a wave momentum flux $M = E / c_p$, where $E \propto a^2$ is the mean wave-energy density, reduction in the wave amplitude implies a reduction of the wave momentum flux. Conservation of momentum then induces a current in the water under the ice, and / or forward stress in the ice cover. A viscous, second order, weakly non linear derivation of the induced currents under a thin cover with longitudinal elasticity is presented in \citet{christensen2005transient}. While the derivation was initially intended to describe the flow under a thin surfactant film, the equations present some similarities with what would be obtained due to a thin elastic cover with negligible rigidity and predicts the existence of modified currents (relatively to the Stokes drift observed in the free water cases) under surface covers. However, wave momentum flux diminution is not the only effect that plays a role in the development of currents under the ice. In particular, in the case of grease ice, \citet{Martin1981} observed a mean current in the opposite direction to what would be expect from wave momentum flux considerations. This current was found to be a side-effect of mass conservation, when the grease ice layer gets pushed alongside the wave tank by the incoming waves.

\section{Main axis of the PhD work}

The initial objectives of the present PhD thesis were to further investigate wave propagation and damping in sea ice through a comparison of field and laboratory results, and to study associated induced currents that could participate in the dispersion of pollutants in the Arctic regions.

Regarding the first two objectives, the need for further investigation of wave propagation and damping is made clear by the absence of consensus in the literature regarding which wave damping model to use \citep{JGRC:JGRC21350}. Indeed, the debates taking place can partly be explained by the limited amount of wave data available, relatively to the wide range of ice conditions that can be encountered. More field and laboratory data are necessary to both calibrate models, and offer an exhaustive overview of the effect of different sea ice conditions on wave propagation. Therefore, a large part of our effort has been to perform new measurements, and to participate in the development of instruments and methodology for performing accurate field and laboratory measurements of waves in ice.

Finally, the presence of wave-induced currents under the ice is a topic that, as pointed out in the literature review, has been little investigated in the past. Regarding this last point, I focused on performing measurements using Particle Image Velocimetry (PIV) and Particle Tracking Velocimetry (PTV) in several cases representative of different ice conditions.

More details about both the field and laboratory work performed are available in the next sections.

\subsection{Fieldwork investigation of waves in ice}

Fieldwork investigations performed in this project have focused on measuring wave-ice interaction in the region around Longyearbyen, Svalbard during the winters 2015, 2016, and 2017 \citep{7513396, JGRC:JGRC21649, RabaultSutherlandGlaciology}, and in the Barents sea during the spring 2016 \citep{marchenko2017field}. Various types of ice were observed during those measurements, including land-fast ice, grease ice, large ice floes and broken ice floes. As indicated in the previous sections, there is a critical need for more field measurements to be performed at a reduced price, in order to quantify the sea ice dynamics happening in the Arctic.

The first difficulty when performing measurements in the arctic is set by the harsh environmental conditions in which instruments must operate \citep{10.1109/MPRV.2010.53}. As a consequence, commercial solutions are expensive and the cost implied by the deployment of a large number of commercial instruments would have been a challenge for the project. Therefore, instruments built in-house based on Inertial Motion Units (IMUs) were developed for performing measurements of waves in ice. Those developments started with the use of Moxa industrial computers during the first measurement campaign, before shifting to (more adapted) microcontroller-based loggers. While the number of instruments built was limited, and, therefore, the hardware is the result of handcrafting and prototyping rather than a commercial product, I took the decision to release all code and designs as open source material. This presents several advantages. In particular, sharing all the details of the design of an instrument can make it easier to reproduce experiments, by drastically reducing the cost and time necessary to build an exact copy of the instrument initially used. In addition, this makes it easier for different groups to build upon a common platform, therefore, encouraging modularity and reuse of previous designs rather than fragmented in-house development, which very likely is redundant between research groups and private suppliers, leading to unnecessary costs. Finally, resorting to open source instruments gives access to the details of their internal functioning, therefore, allowing better transparency over how data is collected and pre-processed compared with many commercial solutions.

The data obtained from the IMU instruments can be used in several ways. The most common uses for the data, which are presented in our publications \citep{JGRC:JGRC21649, RabaultSutherlandGlaciology, doi:10.1175/JTECH-D-16-0219.1}, include computing the wave energy spectrum (and, from it, other important quantities such as the significant wave height, peak wave frequency, and other spectral characteristics), and estimating the wavelength and dispersion relation.

Once the power spectrum density of acceleration data $S[\eta_{tt}]$ has been computed, usually by means of the Welch method to reduce the noise level, one must convert it to obtain the power spectrum density $S[\eta]$ of wave elevation, following \citet{tucker2001waves}:

\begin{equation}
  S[\eta] = \omega^{-4} S[\eta_{tt}].
\end{equation}

The wave spectrum $S$ gives information about the wave energy content at different frequencies. It can be used in a variety of ways, including most interestingly for waves in ice to compute the wave damping if several measurements are performed at different points in space. Considering $\alpha(f)$ the frequency dependent spatial damping coefficient, $a_{i/j}(f) = \sqrt{S_i(f) / S_j(f)}$ the attenuation coefficient for waves propagating from sensor $j$ to sensor $i$, and $d_{i, j}$ the distance along wave propagation direction between those sensors, the damping can be obtained as:

\begin{equation}
  \alpha(f) = - \frac{\log(a_{i/j}(f))}{d_{i, j}}.
\end{equation}

Wavelength can also be obtained from comparing the phase shift between sensors separated by a known distance along the direction of wave propagation. Indeed, the phase difference between sensors $i$ and $j$ can be expressed both from the definition of the wave number:

\begin{equation}
  \phi_{i, j}(f) = k(f) d_{i, j},
\end{equation}

\noindent as well as from the co-spectral density $S_{i, j}(f)$ between sensors $i$ and $j$:

\begin{equation}
  \phi_{i, j}(f) = \tan^{-1} \left( \frac{Im[S_{i, j}(f)]}{Re[S_{i, j}(f)]} \right),
  \label{Eqn:cospectral}
\end{equation}

\noindent where $Im$ and $Re$ indicate the real and imaginary part, respectively. To get best results, the distance between the sensors used for performing a cross correlation analysis should be typically half a wavelength. For distances larger than a wavelength, signal aliasing complicates the processing.

I pushed this approach further in \citet{marchenko2017field} and computed both the dispersion relation and the peak direction of the waves, by using a triad of sensors in a triangle configuration. Each pair of sensors in the triad is used to compute the component of the wave vector along the corresponding direction, so that using three non aligned sensors the full wave vector can be reconstructed from elementary trigonometry.

Following the experience gained from a series of measurement campaigns, I now believe that the configuration used in \citet{marchenko2017field} is the most promising for performing detailed wave measurements in the field using IMU instruments. In this experiment, a series of 3 triads of IMU instruments (plus one individual instrument, as $10$ were available) were positioned over a direction roughly aligned with the direction of wave propagation observed during deployment. As a consequence, both the wave vector (and, hence, dispersion relation) and damping can be obtained in theory. Unfortunately in this experiment, the distance between the sensors (limited by the size of the ice floe they were all deployed on, for logistics reasons) was too little in relation with the wave damping to estimate $\alpha$ with significant error bars. Nonetheless, I hope to be able to deploy again instruments in such a configuration over a larger distance in landfast ice in the future.

Finally, attempts have been made to measure currents in the field using an Acoustic Doppler Velocimeter (ADV) and an Acoustic Doppler Current Profiler (ADCP). No usable data has been generated yet, but it is planned to perform further deployments in the years to come.

\subsection{Laboratory investigation of the interaction between sea ice and waves}

As the goal of this wave in ice study is to describe phenomena happening at larger scales in the nature, field measurements are the final truth against which one should aim to compare models and theory. However, performing experiments in the laboratory presents several advantages compared with field studies:

\begin{itemize}
  \item Laboratory experiments can be performed in a standardized fashion, therefore, allowing a higher reproducibility of results over what can be achieved in the necessarily complex situations encountered in nature. This is especially true in the case of wave / ice interactions, where for example the ice properties and water eddy viscosity have been shown to drastically vary from one measurement to another \citep{DeCarolis2002399, TIMCO2010107}.

  \item Laboratory experiments give us control (to some extent) over which physical phenomena are dominant in a given experiment, and which ones can be neglected. For example, concerns have been raised in the case of field measurements of wave attenuation in the marginal ice zone over how much the wind input to the wave spectrum can influence results \citep{li2017rollover}. This is naturally easily avoided in the laboratory.

  \item Laboratory experiments are much better adapted than the ocean for applying some of the most informative - but also most challenging to deploy - measurement techniques, such as Particle Image Velocimetry (PIV). This allows to obtain information about the underlying physics that could not have been obtained in the field.
\end{itemize}

Therefore, this project has relied on a combination of field and laboratory experiments, in which the phenomena observed in field experiments were reproduced in the laboratory for further investigation. For this, three facilities were used:

\begin{enumerate}
  \item An intermediate size wave tank at the University of Oslo. The wave tank is $24.5$~m long, $0.5$~m wide, and can be filled with up to $0.7$~m of water \citep{Sutherland201788}. It is equipped with an hydraulics-controlled piston type paddle. The water cannot be frozen and, therefore, no real ice can be used. As a consequence, the ice was modeled using either a thin latex sheet (thickness: $0.25$ or $0.5$~mm), or a plastic cover (PEHD, thicknesses: $1$, $3$ and $4$~mm). This wave tank is well equipped, with both ULS ultrasonic gauges (resolution $0.18$~mm), a white LED lightsheet, and a Falcon 2 camera that can be used to record PIV images. This facility was used for the work presented in the following papers: \citet{Sutherland201788, rabault2016ptv}.

  \item The second facility is a small scale wave tank ($3.5 \times 0.3$~m, filled with up to $0.25$~m of water), located in a cold laboratory in Longyearbyen, Svalbard. A piston paddle was built for that tank, and an open source control system was designed (for more information, see the Appendix A). Both Banner ultrasonic gauges (model S18U, resolution $0.5$~mm) and the same PIV setup as in Oslo were used in the cold room. As the tank is entirely made of Plexiglas, easy optical access to the water and the side of the ice are available. When growing ice, isolation foam is put around the tank to make sure that ice grows from the surface down into the water. This tank was used for the work presented in \citet{rabaultgreaseiceSvlbd}.

  \item The last facility is the ice wave tank of HSVA, located in Hamburg, Germany. This is a large climate controlled wave tank, of dimensions $78 \times 10$~m, which can be filled with up to $2.5$~m of water. As optical access is not possible, PIV could not be used. Instruments provided by the university of Oslo during those experiments include a set of Banner ultrasonic gauges, an ADV and an ADCP. As measurements at HSVA were performed towards the end of the project, analysis of the data is still under work.
\end{enumerate}

All measurements in the laboratory presented later in the Thesis were performed using monochromatic waves. In addition, a few measurements resorting to a JONSWAP spectrum were performed at the university of Oslo, but analysis of the results proved challenging. As previously stated, three kinds of measurements were performed, depending on the facility used: measurements of wave elevation using ultrasonic gauges, measurements of the water motion using PIV / PTV, and measurements of the water column velocity profile using ADCPs (ADCPs were only used in the last set of experiments in Hamburg).

Ultrasonic gauges provide high frequency measurements (up to $200$~Hz) of the wave elevation. Calibration is performed either automatically by the gauges (ULS gauges), or manually by both setting the gauges range and measuring the water height in a set of reference cases (Banner gauges). The information obtained from ultrasonic gauges can be used in a very similar way to the IMU data obtained in the field. In particular, the Fourier spectrum of the wave elevation signal, $S(f)$, can be used to obtain the wave amplitude $a$ by integrating around the harmonic frequency:

\begin{equation}
  a(x) = \sqrt{\int_{f_0 - \Delta f}^{f_0 + \Delta f} S(f, x) df},
\end{equation}

\noindent where $f_0$ is the harmonic frequency, $\Delta f$ the half frequency integration width, and $x$ the distance along the wave-tank. A fit of the wave attenuation can be performed on the amplitude data obtained, in a similar way to what is performed with field data.

Information about the wavelength and, therefore, the dispersion relation, can also be obtained from ultrasonic gauges. By setting up gauges in such a way that two of them are separated by typically half a wavelength, one can suppress aliasing effects and obtain a measurement of the wavelength by either a correlation or a cross-spectrum analysis, in a similar way to what is presented in Eqn. (\ref{Eqn:cospectral}).

One of the advantages of laboratory measurements is to allow (depending on the construction of the wave tank) easy optical access to the water and, therefore, one can perform direct optical measurements of the water velocity field. Compared with instruments such as ADCPs and ADVs, optical measurements have the advantage of being non intrusive and more accurate. Optical measurements performed in this project relied on a simple mono-camera setup. The light source used was a white light LED array, which is easier and safer to deploy than PIV lasers. A Falcon 2 4M camera was used to acquire images at a frequency of $75$ frames per second.

The optical images recorded were used in two ways. In the case of the experiments performed at the University of Oslo under latex or PEHD covers, the damping is very small and boundary layers are too thin to be visible. Therefore, there is little deviation to see in the instantaneous velocity field, and the work performed focused on studying the steady currents under the surface cover. As a consequence, Particle Tracking Velocimetry (PTV) was used, as it can follow the particle drift without introducing the errors that would be obtained from integrating PIV velocity fields. Obtaining unambiguous results about mean currents under waves is difficult, as a number of second order effects happen simultaneously and cannot be studied separately. These effects include the elasticity induced effect that was the main object of interest and the Stokes drift, but also recirculation in the tank, effects of the start of the wave train, and possibly large convection motions due to temperature differences in the room. Despite these difficulties, it was possible to observe an effect of the surface cover \citep{rabault2016ptv} that is in agreement with the predictions of \citet{christensen2005transient}.

By contrast, the damping was much higher in the experiments performed under grease ice in the wavetank in Svalbard, and rich dynamics could be observed in the water when blocks of grease ice collided. As a consequence, PIV was used in this case. In order to extract information from all the PIV snapshots into a few reduced modes, the Proper Orthogonal Decomposition (POD) relying on the snapshot method was used. The POD is computed from the Singular Value Decomposition (SVD) of the snapshot matrix (for detailed explanations about POD and the snapshot method, see for example \citet{berkooz1993proper, Kerschen2005}). The snapshot matrix, $X$, is constructed as:

\begin{equation}
  X =
  \begin{bmatrix}
    u_1^1       & ... & u_n^1  \\
    \vdots      & ... & \vdots \\
    u_1^k       & ... & u_n^k \\
  \end{bmatrix},
\end{equation}

\noindent where $u_i^j = u(\bm{x}_j, t_i)$, with $\bm{x}_j$ the position of the point considered, and $t_i$ the time of the snapshot. Both the $X$ and $Y$ components of velocity are stored in $u$, by letting the $u_i^{1..k/2}$ represent the $X$ component and the $u_i^{k/2+1..k}$ represent the $Y$ component, respectively. As a consequence, each column of $X$ contains the 2D, 2-components velocity field at a given time reshaped into a 1D vector. The SVD decomposition is then computed as:

\begin{equation}
  X = U S V^*,
\end{equation}

\noindent where $U$ and $V$ are unitary matrices. The diagonal coefficients of $S^2$ give the energy of each mode, while $U$ and $V$ contain the POD modes and POD mode coefficients. The POD modes together with the POD mode coefficients contain the full information about the velocity field at all times. In particular, the description of the velocity field obtained with POD is interesting as it optimizes the energy content in the modes of lower index, therefore, effectively extracting an ordered list of the most energetic coherent structures in the flow.

The results obtained from POD allow for a direct confirmation of the exponential attenuation of the waves under the ice. In addition, a recirculation current similar to the one observed qualitatively by \citet{Martin1981}, is detected. Finally, in the case of colliding grease ice packs, eddy structures are detected. This provides a possible mechanism for the injection of eddy motion in the field.

\section{Unsuccessful directions, future work and a few personal thoughts}

While one usually only reports positive results in the published literature, I find that this Thesis may be the place to also talk about unsuccessful directions and problems encountered in the project.

No real unsuccessful attempts were encountered in the field, but a number of difficulties inherent to working with sea ice restricted the amount of data collected. In particular, the winters 2016 and 2017 were too mild to encounter ice conditions similar to the landfast ice found in 2015, which was the ideal case for studying large continuous ice sheets \citep{JGRC:JGRC21649}. Looking back with the experience acquired during the project, it is clear to me that our group was extremely lucky collecting the 2015 dataset. Indeed, both landfast ice and incoming waves were present, and rich dynamics including the development of cracks in the ice cover were observed. In addition, the direction of the wind kept the ice packed into the fjord instead of scattering it towards the open sea and, therefore, the instruments could be recovered. Had the wind blown towards the open ocean, the instruments (and wave records) would have been lost. This pinpoints the necessity of including some satellite data transmission to our instruments, which our group will be working on in the years to come.

\begin{figure}
  \begin{center}
    \includegraphics[width=.3\textwidth]{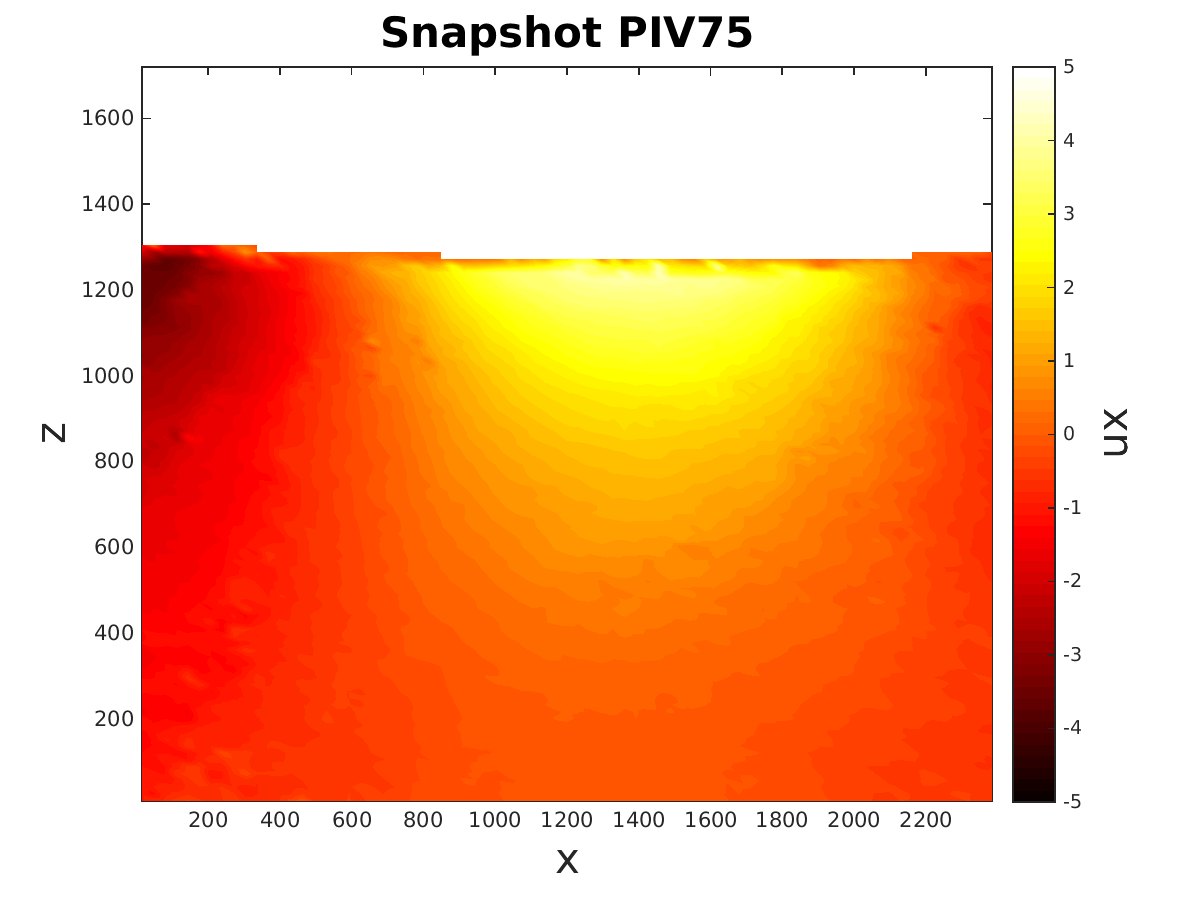}
    \includegraphics[width=.3\textwidth]{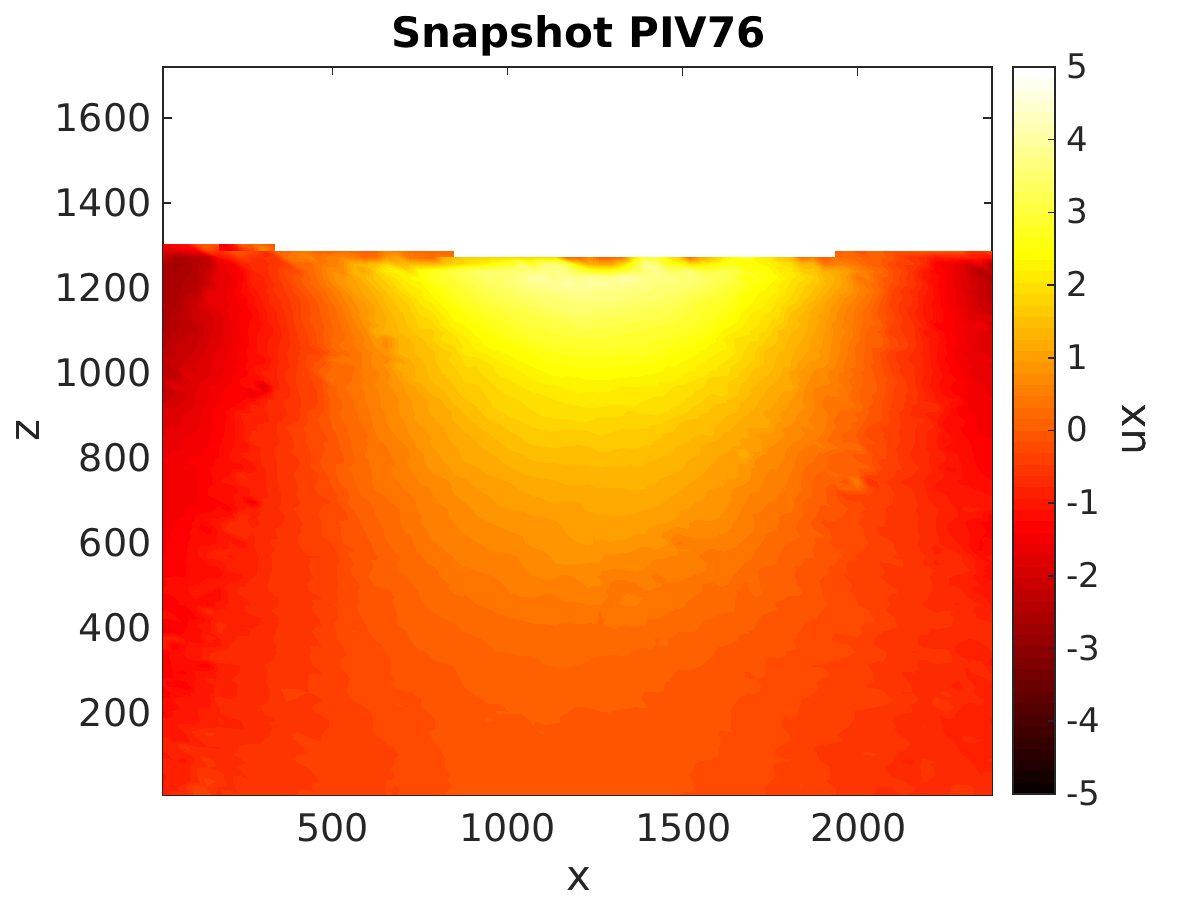}
    \includegraphics[width=.3\textwidth]{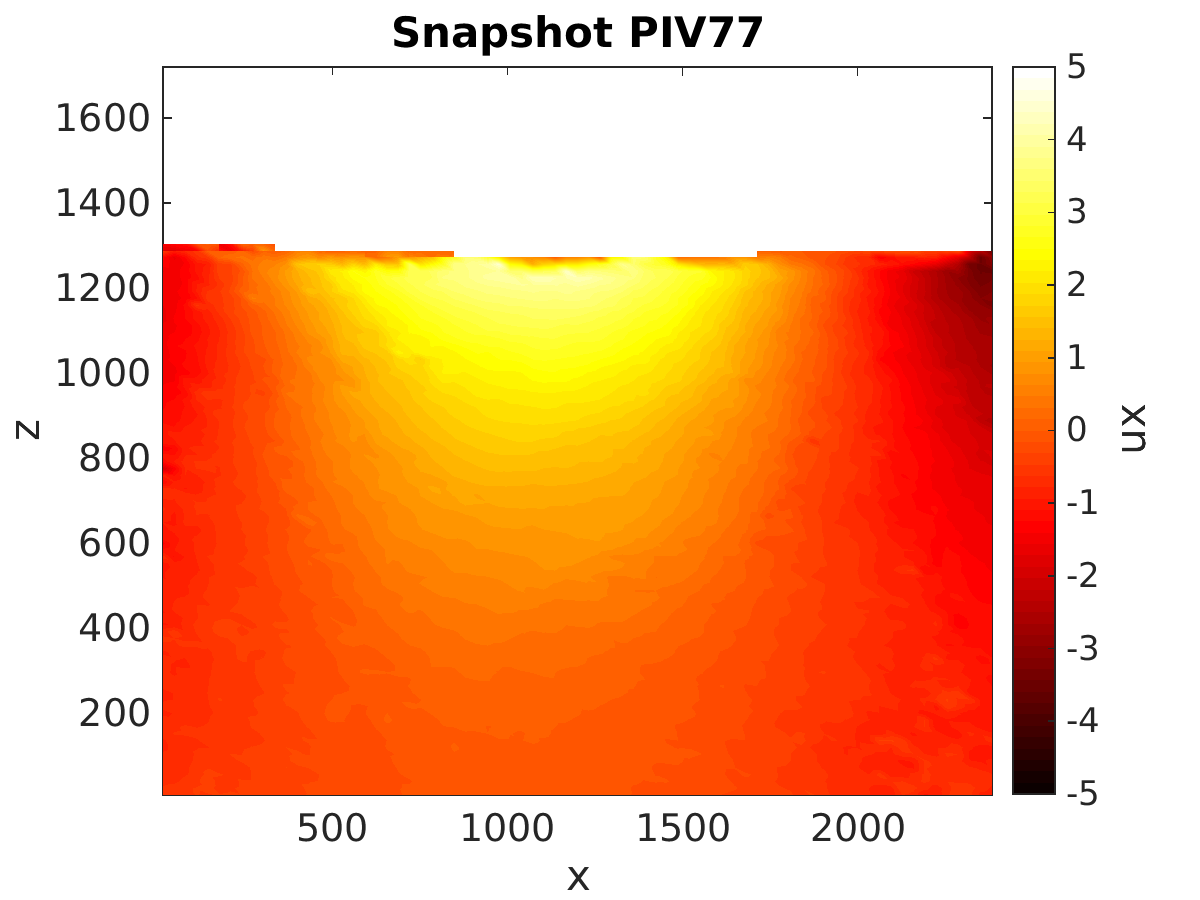} \\
    \includegraphics[width=.3\textwidth]{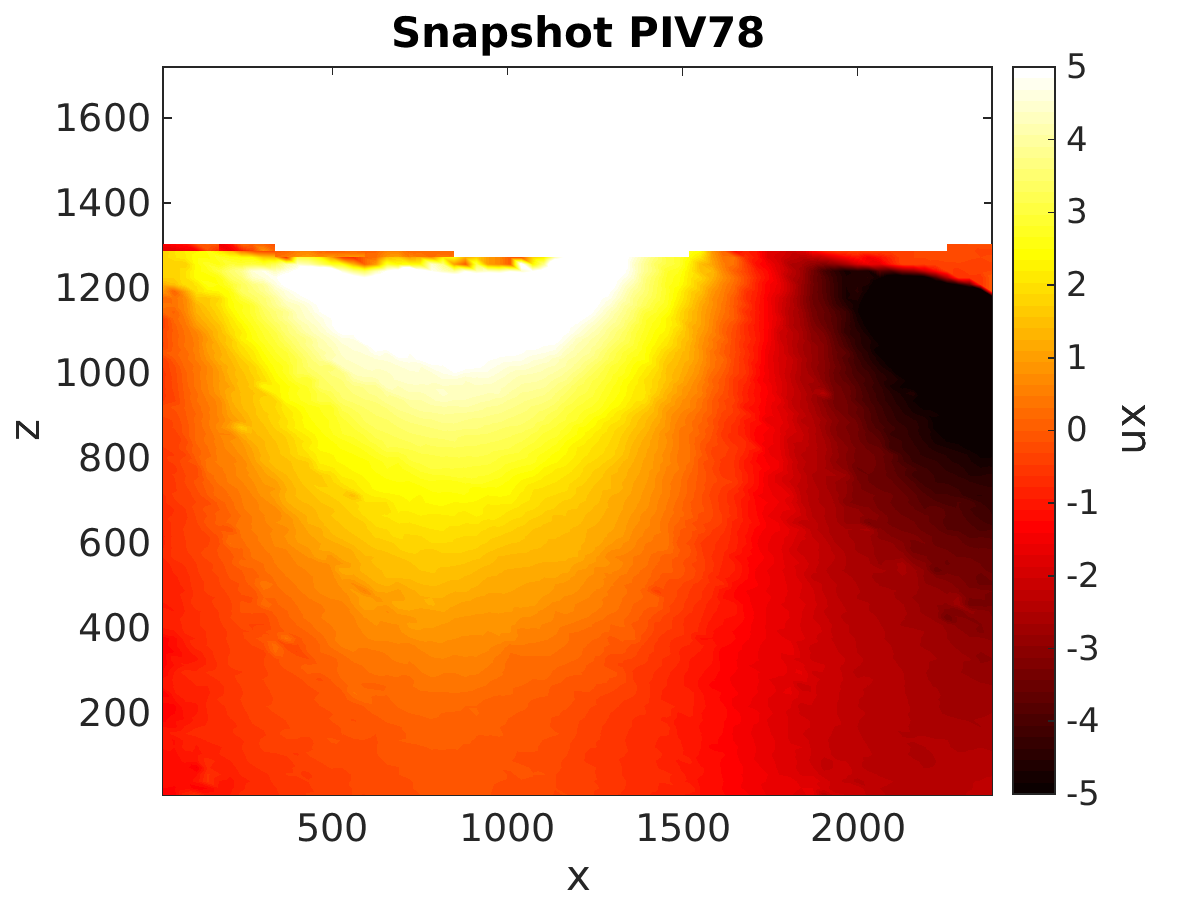}
    \includegraphics[width=.3\textwidth]{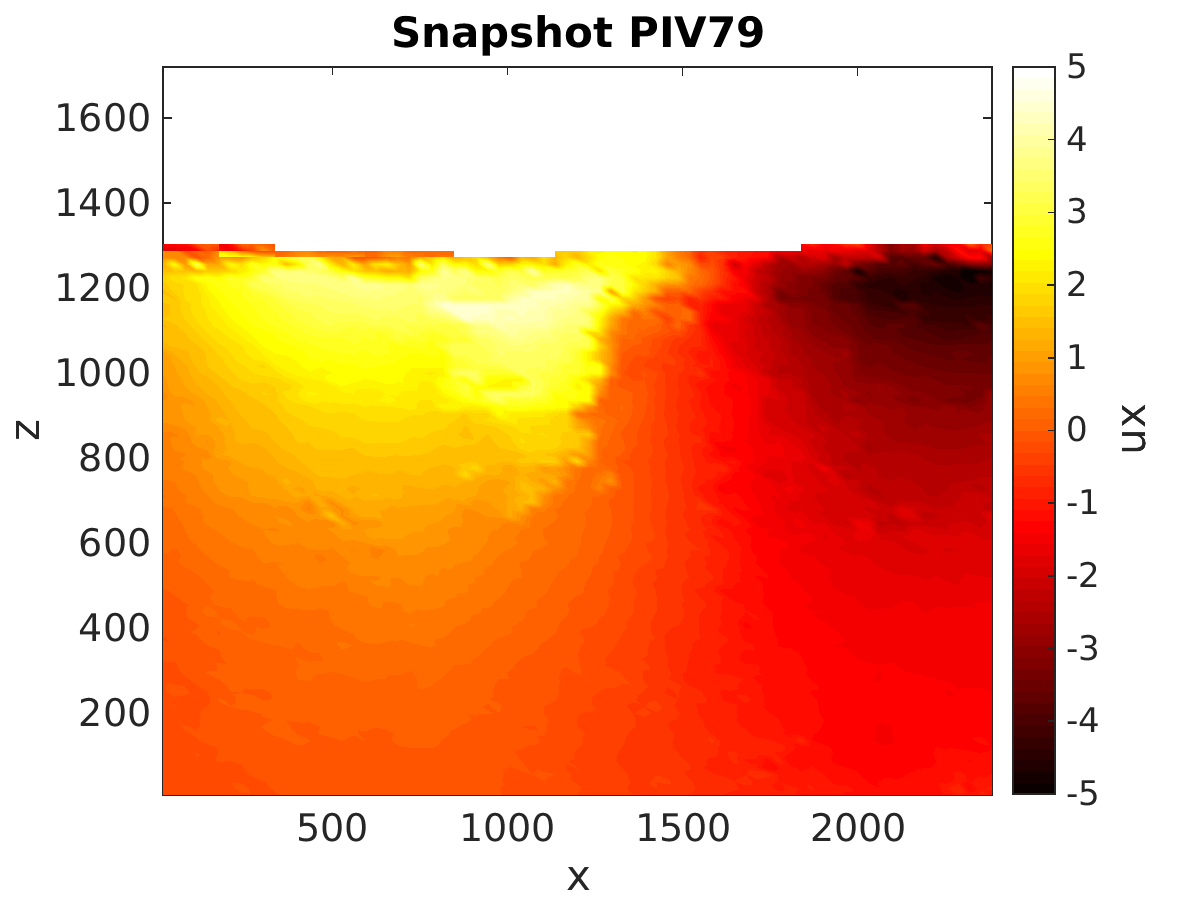}
    \includegraphics[width=.3\textwidth]{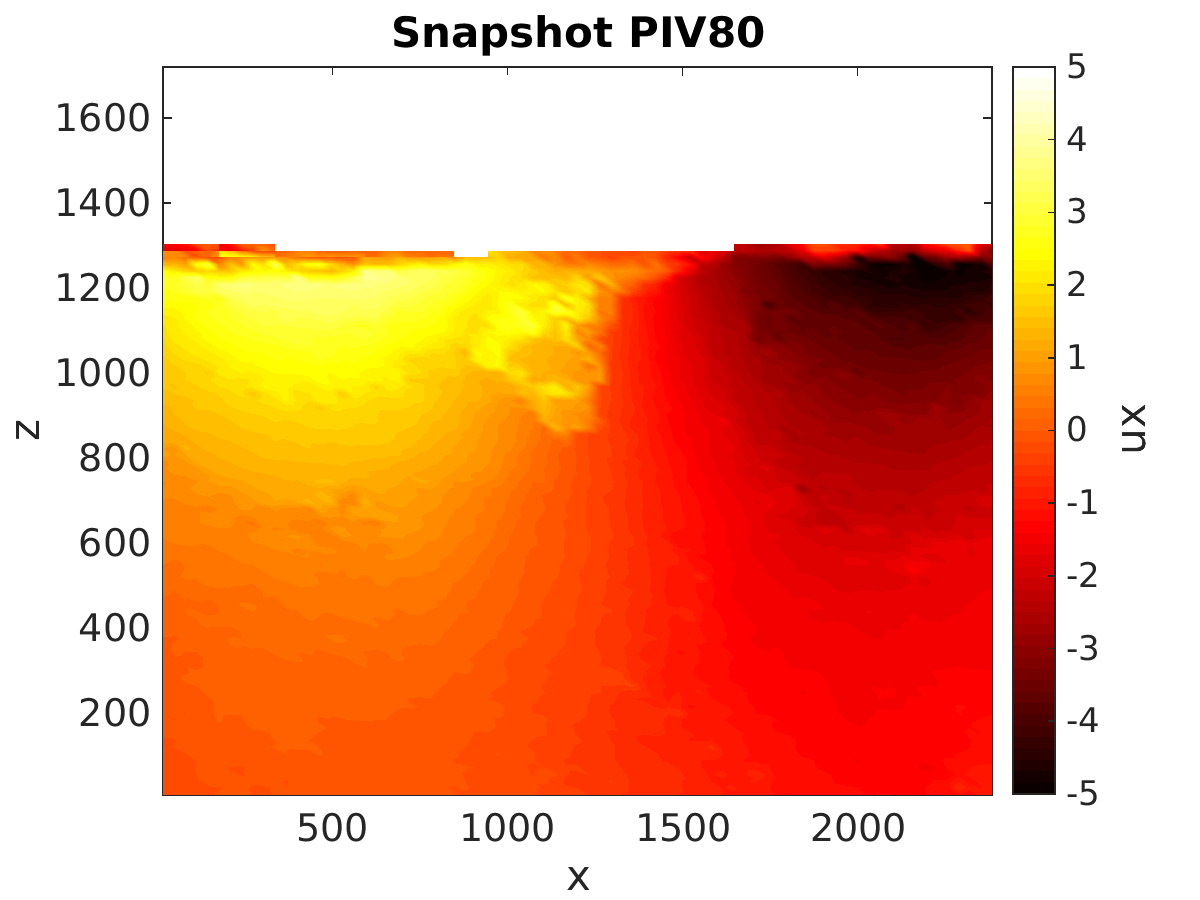} \\
    \includegraphics[width=.3\textwidth]{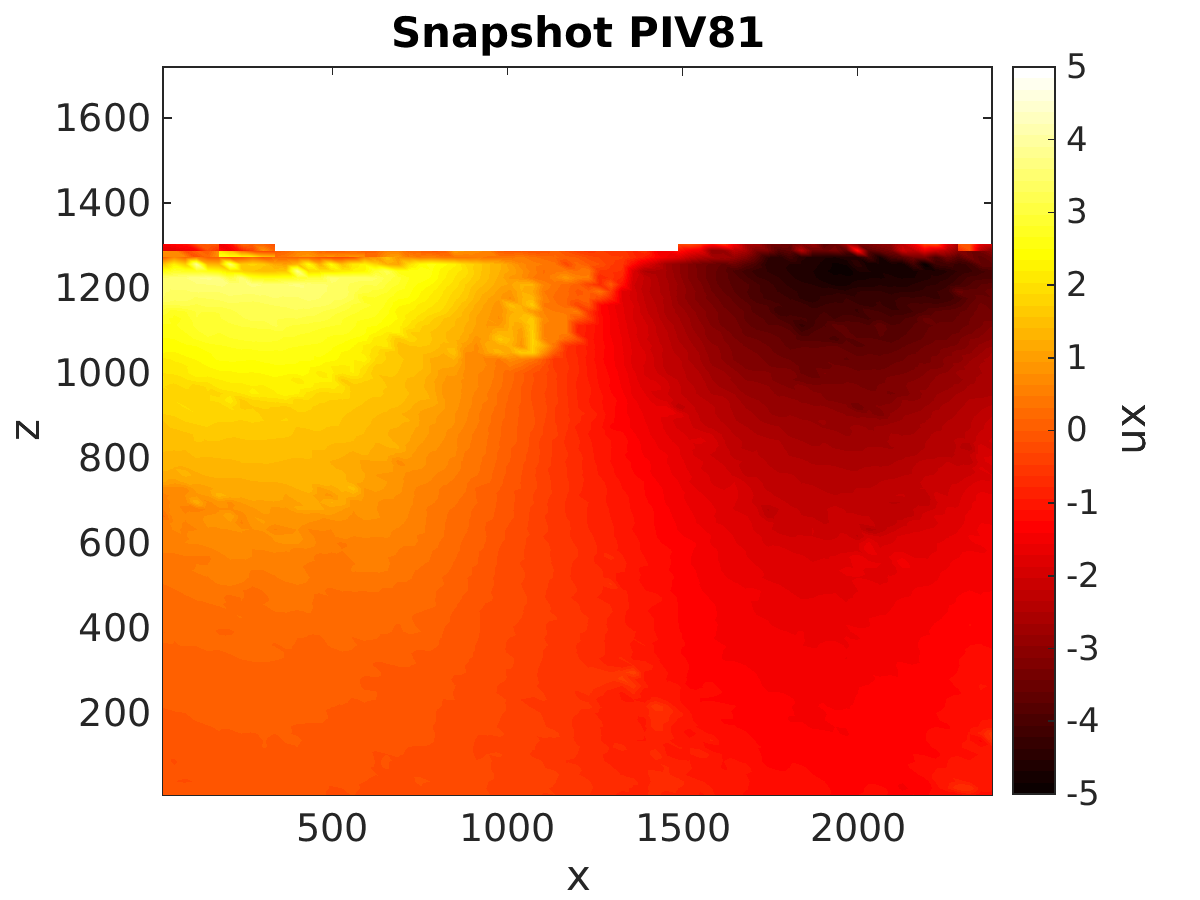}
    \includegraphics[width=.3\textwidth]{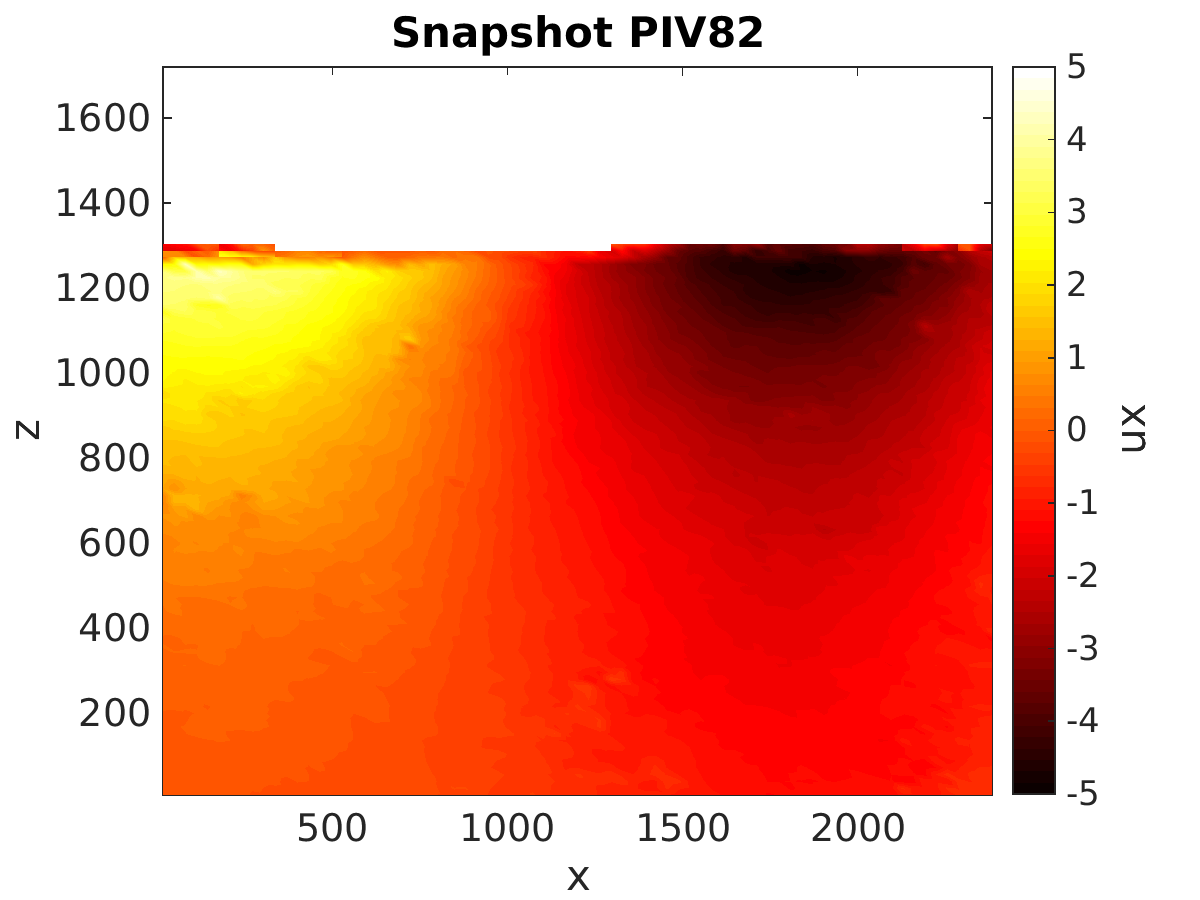}
    \includegraphics[width=.3\textwidth]{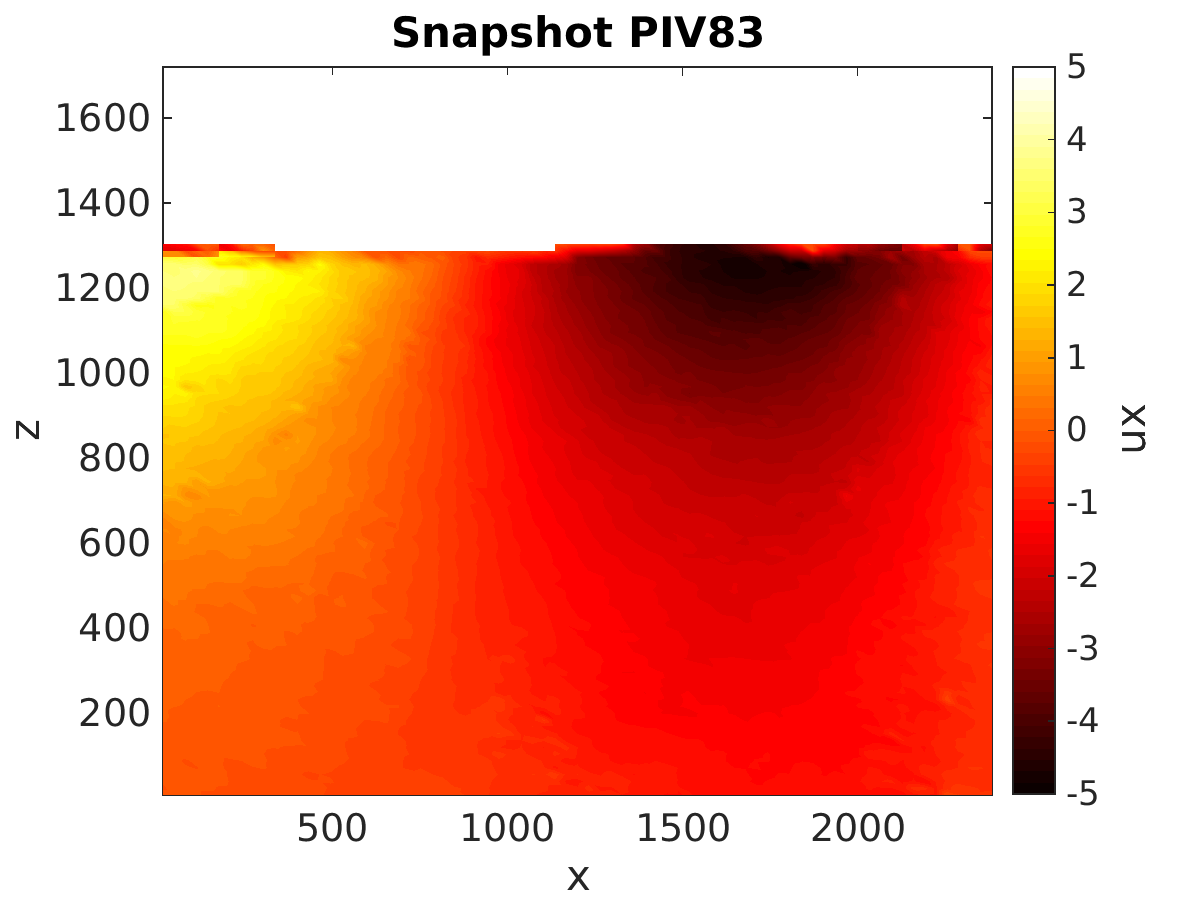} \\
    \caption{\label{fig:turbulent_bursts} Raw PIV snapshots obtained from PIV images taken by Trygve Halsne during the course of his Master Thesis \citep{halsne2015eksperimentell}. It proved impossible to reproduce similar turbulent bursts in subsequent experiments and, therefore, one can expect that an external source of disturbance probably is necessary to trigger the instability.}
  \end{center}
\end{figure}

By contrast, at least one unsuccessful direction was found in the laboratory experiments, which regards the detection of turbulent bursts under waves propagating under inextensible latex covers. In a series of experiments performed during the Master Thesis of Trygve Halsne \citep{halsne2015eksperimentell}, turbulent bursts were observed for high steepness waves (see Fig. \ref{fig:turbulent_bursts}). However, I was not able to reproduce those observations in later experiments, despite several months of investigation at the University of Oslo wave tank. A possibility is that additional disturbances, such as the presence of small bubbles under the cover, or water runup at the leading edge of the latex sheet, helped trigger those bursts in the original experiments and were later removed.

In addition, concerns should be raised about the validation of laboratory experiments compared with field measurements. Indeed, the Reynolds number is not scaled between waves at sea and wave experiments performed in the laboratory. More specifically, the very first imperative of laboratory experiments on waves is to be in the right water depth regime. Indeed, the wave velocity profile and the distribution of wave energy with depth is obviously very different between deep and shallow waves, and can have an effect on the physics at stake if some of the dissipation happens in the water under the ice. Due to the very limited depth of most wave tanks, this puts a sharp requirement on the minimum frequency that can be used for tests in the laboratory. Considering as a first approximation in all the discussion that follows that the deep water dispersion relation is enforced, so that $\omega^2 = gk$, one needs to enforce typically $kH > 1$, i.e. $\omega > \omega_{min} = \sqrt{g/H}$. In addition, the steepness of the waves must be kept moderate to limit nonlinear effects, typically $ \epsilon = k a \approx 0.1$, i.e. $a < 0.1 g / \omega^2$, where $k$ is the wave vector and $a$ the wave amplitude. If one considers that the amplitude-based Reynolds number defined in \citet{GRL:GRL22071}, $Re_a = a^2 \omega / \nu_{water}$, is the right non-dimensional parameter for describing turbulence under waves and large eddy structures under the ice, one then finds that in a laboratory experiment, with a wave-tank of depth H:

\begin{equation}
  Re_a = \frac{a^2 \omega}{\nu_{water}} < \frac{10^{-2} g^2}{\omega^3 \nu_{water}} < \frac{10^{-2} g^{1/2} H^{3/2}}{\nu_{water}}.
\end{equation}

This implies that, even in the case of a wave-tank several meters deep, the maximum Reynolds number that can be achieved in the laboratory for deep water waves of reasonable steepness is much lower than any typical swell on deep water. More specifically, the maximum Reynolds number obtained in a wave tank scales as the water depth to the power $3/2$. As a consequence, studying phenomena related to water wave turbulence and wave-induced eddy viscosity is challenging in the laboratory, as it is impossible to investigate whether a higher Reynolds number could lead to different physics. In particular, the Reynods number attained in our laboratory experiments in a cold room make the full investigation of theories about wave-generated turbulence challenging. One could argue that turbulent flows exhibit a similar behavior above a critical Reynolds number which depends on the flow configuration, and therefore one could expect that reasonable results regarding the effect of turbulent processes could be obtained in large scale laboratory experiments without reproducing the full-scale Reynolds numbers, but more research is needed in this direction before any definitive statement can be made.

On a more technical point of view, this project has underlined the need for further development of a series of affordable, open source instruments relying on GPS, IMUs and Iridium communications for performing measurements of waves in ice. The capability and accuracy of the instruments developed to fulfil these requirements during the project have been carefully evaluated, to make sure that they can be used in a research context. While instruments very similar to those that will be further developed by our group are already used by another group of researchers \citep{doble_mercer_meldrum_peppe_2006, doble2017robust}, relatively little information is available about the details of their design. In the years to come, more work will be performed towards developing general purpose instruments and making their design freely available. This is, in the opinion of the author, the best answer that can be provided to the need of collecting drastically more field data in the polar regions.

\section{Conclusion}

The polar regions are the focus of increased attention due to a combination of environmental, political and economic reasons. One of the many factors that deeply influence these regions is the interaction between waves and ice. Prior to this dissertation, it had become clear that successful modeling of wave-ice interaction is challenging. In particular, both field and laboratory data were still too scarce to unambiguously point towards one successful wave-ice interaction model, and as a consequence a number of competing models exist where each model can be approximately fitted to the observations available. To remedy this problem, new experimental insights are needed.

In the present work, efforts were directed towards helping clarify these questions through two main axis. First, much work was performed towards the development of new methodology and instruments to perform measurements of the ice dynamics in the Arctic. Second, several datasets coming from both field and laboratory experiments were collected, analyzed, and compared with several theories. While more work is needed to bring definite answers, this work has contributed to turning the focus of research at least partly towards the dissipative phenomena happening in the water layer under the ice, rather than only the rheological properties of the ice cover as was implied by some of the recent models for wave propagation in ice covered sea.

Building, calibrating and validating models that integrate the effect of both the rheology of the ice and the dynamics of the water beneath the ice is made difficult by the complexity, both physical and analytical, that is implied by describing all these phenomena at once. However, it appears as one of the long-term goals of wave-ice interaction study. Much work probably remains before such a complete model can be developed into maturity, and a number of complex phenomena that were not discussed here will also probably play a role, such as ice breakup by wave forcing.

\section{Appendix: some technical aspects: an overview of the open source tools released in the course of the project}

Several designs and code bases have been released as open source in the course of this project, both as code on the Github of the author and as a series of web pages. A list of released materials is provided here:

\begin{itemize}
  \item The code and general design of the instruments that performed measurements of waves in ice using IMUs are presented in \citet{RabaultSutherlandGlaciology} and released on github:
  \begin{center} https://github.com/jerabaul29/LoggerWavesInIce . \end{center}
  In addition, a web page presents some technical aspects of the systems used:
  \begin{center} https://folk.uio.no/jeanra/Microelectronics/MicrocontrollerBasedLoggers.html . \end{center}

  \item The paddle control system and the coupled logging system used for the wave tank measurements in Svalbard are released on github:
  \begin{center} https://github.com/jerabaul29/ArduinoPaddleControl \end{center}
  \begin{center} https://github.com/jerabaul29/PaddleAndUltrasonicGauges . \end{center}
  This system allows to perform automated measurements for a range of wave amplitude and frequencies.

  \item The logging system used during the experiments at HSVA is released on github:
    \begin{center} https://github.com/jerabaul29/logging_ultrasonic_gauges . \end{center}
\end{itemize}

\chapter{Selected publications}

This chapter is a selection of 7 works that were (or are being) published during the course of the project. Those publications are chosen to reflect both the laboratory experiments and fieldwork performed.

\begin{enumerate}
  \item The first publication (Sutherland, G. and Rabault, J. (2016), "Observations of wave dispersion and attenuation in landfast ice", {\em Journal of Geophysical Research: Oceans}, 121(3):1984--1997) is an analysis of the measurements performed during the first measurement campaign in Svalbard. It includes discussions about wave propagation and attenuation. It is one of the few studies published in the literature where both the wave attenuation and dispersion relation are obtained in the case of field measurements. In this paper, we present a direct measurement of the dispersion relation for waves in continuous ice, and we observe the development of cracks which correspond to a suppression of the flexural effects. \\

  \item The second publication (Rabault, J., Sutherland, G., Gundersen, O., and Jensen, A. (2017), "Measurements of wave damping by a grease ice slick in Svalbard using off-the-shelf sensors and open-source electronics", {\em Journal of Glaciology}, page 1–10) focuses on measurements of wave attenuation in grease ice, obtained during the second measurement campaign in Svalbard. Those data are especially relevant for calibration of wave in ice models. In addition, discussions of a more technical order about both the necessity of developing open source instruments and the importance of quantifying the quality of models using numerical indicators rather than visual observations of quality of fit are presented. \\

  \item The third publication (Sutherland, G., Rabault, J., and Jensen, A. (2017), "A Method to Estimate Reflection and Directional Spread Using Rotary Spectra from Accelerometers on Large Ice Floes", {\em Journal of Atmospheric and Oceanic Technology}, 34(5):1125--1137) is a more in-depth analysis of the rotary spectrum method used in the first publication, as well as a discussion of some difficulties encountered when performing measurements with IMUs. In particular, it is shown that measurements of reflected wave energy and directional spread of the incoming waves can be obtained from a study of the horizontal components of the ice acceleration. \\

  \item The fourth publication (Marchenko, A., Rabault, J., Sutherland, G., Collins, C. O.~I., Wadhams, P., and Chumakov, M. (2017), "Field observations and preliminary investigations of a wave event in solid drift ice in the Barents Sea", In {\em 24th International Conference on Port and Ocean Engineering under Arctic Conditions}) summarizes a series of measurements performed in the Barents sea. There are two notable points with this study. First, several kinds of instruments relying on different measurement techniques (pressure sensors, ADVs and IMUs) were deployed at the same location. The results obtained from all instruments are very similar, therefore, providing a validation of the methodology used. In addition, the deployment configuration of the IMU instruments as a succession of triads presented in this article is a promising way to measure both the wave attenuation and the dispersion relation, without the need for any assumption on the wave direction of propagation. \\

  \item The fifth publication (Rabault, J., Halsne, T., Sutherland, G., and Jensen, A. (2016), "PTV investigation of the mean drift currents under water waves", In {\em Proceedings of the 18th Int. Lisb. Symp.}) is a proceeding article, in which currents under inextensible covers are studied by means of PTV. We reached the conclusion that diffusion of vorticity due to the presence of an inextensible cover does influence the mean currents. \\

  \item The sixth publication (Sutherland, G., Halsne, T., Rabault, J., and Jensen, A. (2017), "The attenuation of monochromatic surface waves due to the presence of an inextensible cover", {\em Wave Motion}, 68:88 -- 96) is another analysis of the effect of thin inextensible covers on wave propagation in the wave-tank, this time focusing on wave damping. It is shown that, in a well controlled wave tank situation where the surface cover is a smooth, non rigid surface, the damping predicted by Lamb is observed. \\

  \item Finally, the seventh work (Rabault, J., Sutherland, G., Jensen, A., Christensen, K.~H., and Marchenko, A. (2018), "Experiments on wave propagation in grease ice: combined wave gauges and PIV measurements", {\em under review}) presents the results obtained in a wave tank experiment in Svalbard, studying wave propagation under a layer of grease ice. Both wave attenuation and wavelength are compared with a series of one layer models. Satisfactory agreement is found between these one layer models and our experimental results. In addition, PIV measurements are performed for the first time in the case of waves propagating under grease ice. The exponential wave attenuation predicted theoretically is well visible, and vorticity created by colliding blocks of grease ice as well as mean currents are observed. \\
\end{enumerate}

In addition, two publications written in the course of the project are not included in this Thesis. This includes one article about measurements of waves in landfast ice (Rabault, J.; Sutherland, G.; Ward, B.; Christensen, K.H.; Halsne, T.; Jensen, A. (2016), "Measurements of Waves in Landfast Ice Using Inertial Motion Units", {\em IEEE Transactions on Geoscience and Remote Sensing}, vol. 54, no. 11, pp. 6399-6408), which is mostly technical and has been extended and completed by Publication 1 and Publication 3, and one work that is the result of a side project (Rabault, J.; Kolaas, J.; Jensen, A. (2017), "Performing particle image velocimetry using artificial neural networks: a proof-of-concept", {\em Measurement Science and Technology} 28 (12), 125301).

\newpage

\section{Publication 1: Observations of wave dispersion and attenuation in landfast ice}
Sutherland, G. and Rabault, J. (2016), "Observations of wave dispersion and attenuation in landfast ice", {\em Journal of Geophysical Research: Oceans}, 121(3):1984--1997.

\section{Publication 2: Measurements of wave damping by a grease ice slick in Svalbard using off-the-shelf sensors and open-source electronics}
Rabault, J., Sutherland, G., Gundersen, O., and Jensen, A. (2017), "Measurements of wave damping by a grease ice slick in Svalbard using off-the-shelf sensors and open-source electronics", {\em Journal of Glaciology}, page 1–10.

\section{Publication 3: A Method to Estimate Reflection and Directional Spread Using Rotary Spectra from Accelerometers on Large Ice Floes}
Sutherland, G., Rabault, J., and Jensen, A. (2017), "A Method to Estimate Reflection and Directional Spread Using Rotary Spectra from Accelerometers on Large Ice Floes", {\em Journal of Atmospheric and Oceanic Technology}, 34(5):1125--1137.

\section{Publication 4: Field observations and preliminary investigations of a wave event in solid drift ice in the Barents Sea}
Marchenko, A., Rabault, J., Sutherland, G., Collins, C. O.~I., Wadhams, P., and Chumakov, M. (2017), "Field observations and preliminary investigations of a wave event in solid drift ice in the Barents Sea", In {\em 24th International Conference on Port and Ocean Engineering under Arctic Conditions}.

\section{Publication 5: PTV investigation of the mean drift currents under water waves}
Rabault, J., Halsne, T., Sutherland, G., and Jensen, A. (2016), "PTV investigation of the mean drift currents under water waves", In {\em Proceedings of the 18th Int. Lisb. Symp.}

\section{Publication 6: The attenuation of monochromatic surface waves due to the presence of an inextensible cover}
Sutherland, G., Halsne, T., Rabault, J., and Jensen, A. (2017), "The attenuation of monochromatic surface waves due to the presence of an inextensible cover", {\em Wave Motion}, 68:88 -- 96.

\section{Work 7: Experiments on wave propagation in grease ice: combined wave gauges and PIV measurements}
Rabault, J., Sutherland, G., Jensen, A., Christensen, K.~H., and Marchenko, A. (2018), "Experiments on wave propagation in grease ice: combined wave gauges and PIV measurements", {\em under review}

\newpage
\begin{small}
\addcontentsline{toc}{section}{~ \\References}
\bibliographystyle{apalike}  
\bibliography{Bibliography.bib}
\end{small}


\end{document}